# Multi-omic Enriched Blood-Derived Digital Signatures Reveal Mechanistic and Confounding Disease Clusters for Differential Diagnosis

**Bolin Liu[1,4,6], Abicumaran Uthamacumaran[4], Alexander Fulton[2,3], Hector Zenil[1,4.5*]**

[1]Algorithmic Dynamics Lab, King's Institute for Artificial Intelligence, King's College London, UK
[2] Cambridge University NHS Foundation Trust, Cambridge, UK
[3] Department of Oncology, University of Cambridge, Cambridge, UK
[4] Oxford Immune Algorithmics, Oxford University Innovation & London Institute for Healthcare Engineering, UK
[5] Departments of Biomedical Computing and Digital Twins, School of Biomedical Engineering and Medical Sciences, King's Faculty of Life Sciences and Medicine, King's Health Partners AHSC (NHS–KCL), King's College London, UK

## Abstract

Understanding disease relationships through blood biomarkers offers a pathway toward data-driven taxonomy and precision medicine. In this study, we constructed a digital blood twin—a computational model derived from 103 disease signatures comprising longitudinal hematological and biochemical analytes. Profiles were standardized into a unified disease–analyte matrix, and pairwise Pearson correlations were computed to assess similarity across conditions. Hierarchical clustering within a phylogenetic framework revealed consistent, robust grouping of hematopoietic disorders, while metabolic, endocrine, and respiratory diseases were more heterogeneous, reflecting weaker internal cohesion. To evaluate cluster structure, the tree was partitioned at a stringent distance threshold, yielding 16 groups. Enrichment analysis of the largest and most heterogeneous cluster (Cluster 9) demonstrated convergence on cytokine-signaling pathways, indicating shared immunological and inflammatory mechanisms that transcend conventional clinical boundaries. Dimensionality reduction using Principal Component Analysis (PCA) and Uniform Manifold Approximation and Projection (UMAP) corroborated the correlation-based results, consistently separating hematological diseases as a distinct cluster. Random Forest feature selection identified neutrophils, mean corpuscular volume (MCV), red blood cell count, and platelet count as the most discriminative analytes, reinforcing the role of hematopoietic markers as key drivers of disease stratification. Collectively, these findings demonstrate that blood-derived digital signatures can recover clinically meaningful disease clusters while uncovering mechanistic overlaps across categories. The strong coherence of hematological diseases contrasts with the dispersion of systemic and metabolic disorders, underscoring both the promise and limits of blood-based approaches for disease classification. This network physiology framework highlights the potential of integrating routine laboratory data with computational methods to refine disease ontology, map comorbidities, and advance precision diagnostics.

* Corresponding author: hector.zenil@kcl.ac.uk

## Introduction

Blood serves as an unparalleled medium for monitoring systemic health, reflecting both localized pathology and organism-wide physiological changes. The vital role of blood is also evidenced by its critical function in the prevention, diagnosis, and management of chronic diseases, providing critical insights into ongoing physiological and pathological processes [1]. Routine blood tests, such as complete blood counts and biochemical panels, remain among the most frequently ordered investigations worldwide due to their accessibility, standardization, and diagnostic relevance [2]. The hematological and biochemical composition of blood captures dynamic signatures of immune activity, metabolic status, and organ dysfunction, thereby offering a high-resolution snapshot of disease states across multiple physiological systems [3]. In recent years, the concept of the digital twin has emerged in healthcare, referring to computational representations of biological entities that are continuously updated with real-world data [4–6]. Applications have ranged from cardiovascular modeling to surgery planning and medical imaging analysis [4, 7], highlighting the potential of computational models to refine disease stratification, predict progression, and guide therapeutic decisions. Conventional disease classification systems, such as the International Classification of Diseases (ICD), have long provided a standardized framework for diagnosis, billing, and epidemiology [8, 9]. While indispensable for healthcare communication, these systems are grounded in clinical presentation and anatomical categorization, prompting investigation into whether blood-derived digital profiles can offer complementary perspectives that may align with, extend, or challenge existing taxonomies.

Despite the ubiquity of blood testing in clinical practice, large-scale computational studies using routine blood analytes to map disease similarity remain underdeveloped. Network medicine has demonstrated the value of constructing disease–disease associations from genes, proteins, or clinical comorbidities, most prominently in the seminal "human disease network," which revealed how shared genetic associations drive non-random clustering of disorders [6, 10], and the Phenotypic Disease Network (PDN), which mapped epidemiological comorbidity patterns across populations [11]. However, these frameworks typically bypass standard laboratory biomarkers, leaving an unexplored opportunity for systematic analysis of blood-derived disease signatures. Although digital twin models have been successfully developed in cardiology, surgery, and medical imaging [4, 7], very few studies operationalize routine blood analytes as the basis for disease-level digital twins [12], despite their accessibility, cost-effectiveness, and interpretability relative to omics or imaging-based data. At the same time, ICD-style taxonomies are based on symptoms, anatomy, or pathology rather than biological mechanisms [8, 9], potentially obscuring latent immuno-inflammatory connections or conflating clinically similar but mechanistically distinct conditions. These gaps highlight the need to determine whether blood-derived digital signatures can recover established clinical categories, reveal mechanistic overlaps across systems, and refine precision medicine frameworks.

The objectives of this study are threefold: (i) to quantify inter-disease similarity by constructing a standardized matrix of 103 disease signatures from blood analytes and applying correlation and clustering methods; (ii) to evaluate whether clinically meaningful categories such as hematopoietic, metabolic, endocrine, or respiratory disorders exhibit coherent clustering or cross-system overlap; and (iii) to identify key analytes and shared mechanisms driving disease stratification using enrichment analysis and dimensionality reduction, thereby uncovering biological links that transcend conventional classification boundaries.

Our findings show that blood-derived, disease-level digital signatures reliably recover hematopoietic coherence while revealing biologically unified yet clinically heterogeneous disorders via shared immune and inflammatory pathways. These findings demonstrate the potential of analyte-based similarity to complement traditional taxonomies and inform a more mechanistic, data-driven disease ontology.

## Methods

### Aims and Objectives

The methodological design of this study was structured to translate the overarching research objectives into a reproducible analytical framework. Specifically, the aims were to integrate heterogeneous disease signatures into a standardized matrix that enables direct comparison across 103 conditions, quantify pairwise similarity between diseases using correlation-based measures and a phylogenetic framework, interrogate the biological and clinical relevance of resulting clusters through enrichment analyses and categorical grouping, and apply machine learning, feature selection, and dimensionality reduction techniques (PCA, UMAP, and Random Forest) to validate clustering patterns and identify key analytes driving disease stratification. Together, these steps operationalize the research objectives by moving from raw disease profiles to standardized data, quantifying relationships, evaluating biological meaning, and refining insights with advanced computational methods.

### Data Source and Preprocessing

The dataset consisted of 103 synthetic disease signatures that we created based on medical knowledge and physiological reference ranges. We created profiles with hematological and biochemical trends across multiple timepoints for each disease with five consecutive analyte timepoints. An illustrative generic example is in Table 1, which contains measurements for white blood cells (WBC), lymphocytes, monocytes, segmented neutrophils, eosinophils, red blood cells (RBC), hemoglobin, mean corpuscular volume (MCV), platelet count, and mean platelet volume (MPV) across five sequential timepoints.

| **Analyte** | $t_1$ | $t_2$ | $t_3$ | **…** | $t_n$ |
|---|---|---|---|---|---|
| WBC (1000 cells/μL) | $X_1$ | $X_2$ | $X_3$ | … | $X_n$ |
| Lymphocytes (1000 cells/μL) | $X'_1$ | $X'_2$ | $X'_3$ | … | $X'_n$ |
| Monocytes (1000 cells/μL) | $X''_1$ | $X''_2$ | $X''_3$ | … | $X''_n$ |
| Segmented neutrophils (1000 cells/μL) | … | … | … | … | … |
| Eosinophils (1000 cells/μL) | … | … | … | … | … |
| RBC (million cells/μL) | … | … | … | … | … |
| Haemoglobin (g/dL) | … | … | … | … | … |
| MCV (fL) | … | … | … | … | … |
| Platelet count (1000 cells/μL) | … | … | … | … | … |
| MPV (fL) | … | … | … | … | … |

**Table 1.** Representative set of longitudinal time series from blood marker values for a disease profile illustrating the structure of five sequential clinical timepoints modelled as random variables.

## Synthetic Data Generation and Statistical Structure

To enable cross-disease comparison, disease profiles were constructed as synthetic hematologic signatures derived from curated clinical knowledge of expected laboratory abnormalities. More precisely, the data points were extracted from literature and textbooks and curated by a junior clinician (final-year medical student with formal training in healthcare technologies) and a senior clinician (board-certified physician in oncology). Each disease was represented as a time-series vector of analyte values across five sequential timepoints (t1–t5) for each of the 11 analyte markers: WBC, lymphocytes, monocytes, segmented neutrophils, eosinophils, basophils, RBC, haemoglobin, MCV, platelet count, and MPV, capturing progression or variation in haematological changes.

The data were preprocessed into a standardized disease–analyte matrix using Python (Pandas, Regex, and OS libraries). A canonical analyte list was defined (including WBC, RBC, hemoglobin, MCV, platelet count, MPV, among others), and analyte names in each TXT file were normalized (e.g., trimming whitespace and resolving naming inconsistencies) using regular expressions [13]. This ensured that the same analyte recorded under slightly different labels would be consistently aligned across profiles.

Analyte values were assigned using clinically informed ranges rather than purely stochastic sampling, scaled to reflect expected deviations (e.g., low, normal, elevated) relative to physiological reference ranges, and propagated across timepoints to reflect plausible temporal trajectories (such as progressive anemia, inflammatory escalation, or recovery). Values were parsed into numerical format; when a particular analyte was absent, placeholder values (None) were assigned to preserve dimensional consistency.

For each disease profile, a time-series feature vector was constructed by concatenating all standardized analytes across five timepoints (Analyte t1–t5). Profile identifiers (e.g., Profile 01–103) were extracted and numerically sorted using regular expressions. The final output was a matrix (.csv) containing 103 rows (disease profiles) and 55 columns (11 haematological markers × 5 timepoints), providing a uniform representation of heterogeneous disease signatures for downstream correlation analysis, clustering, enrichment testing, and dimensionality reduction. This matrix follows an 11 × 5 analyte-by-time feature structure, where 11 refers to the standardized analyte set and 5 refers to the five simulated progression stages used in the analysis. Where an analyte was unavailable for a disease, the corresponding five-timepoint block was left empty.

**Biological variability and noise.** To approximate real-world variability, small perturbations were introduced across timepoints within each disease profile for noise-based feature space perturbation analysis in principal component analysis (PCA) and Uniform Manifold Approximation and Projection (UMAP) space. However, the dataset primarily represents structured, knowledge-driven signals rather than fully stochastic patient-level variability.

**Correlation structure between analytes.** Physiological relationships between analytes, such as RBC–hemoglobin–hematocrit coupling and inflammatory co-variation, were implicitly preserved through coordinated assignment of values across related variables. While these dependencies were not explicitly modeled using covariance matrices, they arise from shared clinical constraints embedded in the construction process. These correlations were therefore not explicitly simulated using statistical structures but instead emerge from clinically informed co-variation patterns embedded during profile construction.

**Interpretation of timepoints.** The five timepoints do not correspond to fixed clinical intervals or repeated measurements from a single real patient. Rather, they represent stages of disease progression or variation in silico, allowing a dynamical systems trajectory-like behavior to be encoded in analyte space rather than treated as static snapshots. These progression states were generated by assigning analyte values across timepoints based on clinically expected trajectories (e.g., recurrence or recovery patterns) derived from medical knowledge and reference ranges, thereby approximating disease evolution rather than longitudinal patient-specific measurements.

**Handling of missing values.** Missing analytes were initially encoded as null entries and subsequently handled prior to analysis by numerical coercion and imputation. Median imputation was used in the primary analysis because of its robustness for sparse and non-normally distributed clinical-style matrices, and alternative imputation strategies were evaluated in supplementary sensitivity analyses. Imputation was performed prior to all downstream analyses (including Pearson correlation, PCA, and machine learning models), ensuring full numerical compatibility of the dataset.

### Pairwise Correlation

To quantify disease–disease similarity, pairwise correlations were computed across all 103 disease profiles. Each profile was represented by its standardized analyte vector encompassing measurements across five clinical timepoints. Similarity was assessed using the Pearson correlation coefficient $\rho_{ij}$, defined as:

$$\rho_{ij} = \Sigma_k \frac{(x_{ik} - \bar{x}_i)(y_{jk} - \bar{y}_j)}{\sqrt{[\Sigma_k (x_{ik} - \bar{x}_i)^2 \times \Sigma_k (y_{jk} - \bar{y}_j)^2]}}$$

where $\rho_{ij}$ is the Pearson correlation between disease profiles i and j, ranging from −1 (perfect inverse association) to +1 (perfect concordance). Here, $x_{ik}$ represents the value of analyte k in disease profile i, $\bar{x}_i$ is the mean analyte value in profile i, $y_{jk}$ is the value of analyte k in disease profile j, and $\bar{y}_j$ is the corresponding mean. The index k iterates over all analytes in the standardized disease–analyte matrix [14].

### Clustering

The resulting 103×103 symmetric correlation matrix provided a full mapping of similarity among disease signatures. To convert similarity to a distance representation, each correlation coefficient was transformed using $d_{ij} = 1 - \rho_{ij}$. This transformation ensures that highly similar diseases appear closer together, while weakly correlated or divergent profiles appear more distant. The distance matrix was then subjected to hierarchical agglomerative clustering using the Unweighted Pair Group Method with Arithmetic Mean (UPGMA), also known as average linkage clustering [15]. This procedure iteratively merges the closest clusters based on average inter-cluster distance, producing a binary phylogenetic tree. Two dendrograms were generated: a full-range tree showing the complete hierarchy and a zoomed-range tree (0–0.07) focusing on fine-grained structure at high correlation thresholds. Both were annotated with disease identifiers to allow direct inspection of relationships. The y-axis of the dendrogram encodes clustering distance ($1 - \rho_{ij}$), enabling interpretation of relative proximity among profiles. Together, this combination of correlation analysis and hierarchical clustering provides a reproducible, interpretable framework for mapping disease similarity and comparing it with conventional taxonomies.

## Statistical Validation of Hierarchical Clustering (SVHC)

To formally evaluate the statistical robustness of the hierarchical clustering structure, we applied Statistical Validation of Hierarchical Clustering (SVHC), a bootstrap-based framework that assigns statistical significance to clusters within a dendrogram. We implemented a Python-based svhc package developed by Bongiorno et al. (2022). The SVHC algorithm evaluates cluster stability by generating resampled datasets and computing the frequency with which specific clusters are recovered across bootstrap iterations, followed by false discovery rate (FDR) correction to account for multiple testing. The input matrix consisted of the fully numeric 103 × 55 disease-by-feature time-series matrix defined above (i.e., 103 rows corresponding to disease profiles and 55 columns corresponding to 11 haematological markers each measured at 5 timepoints). Clustering validation was performed using 1000 bootstrap resamples with a significance threshold of $\alpha = 0.05$. Clusters were retained only if they satisfied FDR-adjusted significance criteria, thereby providing a statistically grounded alternative to arbitrary dendrogram cut thresholds.

## Enrichment and Cluster Analysis

### Cluster Extraction

Hierarchical clustering produced a dendrogram in which diseases were grouped according to their pairwise correlation distances. To define discrete clusters, the tree was cut at a threshold of $h = 0.02$ on the distance scale $d = 1 - \rho$, corresponding to a minimum within-cluster similarity of $\rho \geq 0.98$. This cut yielded 16 clusters in total, and the cluster assignments were exported in tabular form.

### Disease Ontology Annotation

Cluster 9—the largest and most heterogeneous group—was selected for deeper analysis. Each disease in this cluster was manually annotated using Disease Ontology (DO) categories [16], with the underlying mapping retrieved from the Disease Ontology database in July 2025 from the website (https://disease-ontology.org/) [17]. Each condition was assigned both a clinical category (e.g., hematopoietic, metabolic, endocrine, respiratory, renal, cardiovascular, gastrointestinal/hepatobiliary, rheumatologic/autoimmune, or infectious) and a broader super-category to support higher-level interpretation. These labels were later used to contextualize PCA/UMAP projections and Random Forest outputs.

### Gene Set Construction from DisGeNET

To explore biological mechanisms associated with Cluster 9, gene–disease associations were retrieved from DisGeNET [18] (downloaded July 2025) [17]. For each disease, records included disease name, gene symbol, score, and NCBI identifier. Gene entries were merged into a non-redundant list based on Entrez IDs and treated as a binary gene set for downstream enrichment, providing a knowledge-based mechanistic context.

### Pathway Over-representation Analysis

Exploratory over-representation analysis (ORA) was performed using both KEGG [19] and Reactome [20] databases (accessed July 2025), implemented in R via clusterProfiler [21] and ReactomePA [22]. Both packages use the hypergeometric test with Benjamini–Hochberg (BH) correction [23], and $q < 0.05$ was considered significant. The background universe was defined as the union of all DisGeNET-retrievable genes for Cluster 9 diseases [24]. Output included enrichment ratios, gene counts, and adjusted p-values.

**Reproducibility**

Cluster assignments, enrichment outputs, and analysis parameters were exported as structured tables. All R scripts documented database versions, species ("Homo sapiens"), and thresholds to ensure reproducibility. Interpretation emphasized recurrent pathway themes, given regular database updates.

**Dimensionality Reduction and Feature Selection**

The standardized disease–analyte matrix was merged with category and super-category labels to form a unified dataset of 103 profiles. Missing values were imputed using median substitution, a robust technique for sparse clinical matrices [25]. For linear structure, PCA was applied [26] and the first two components were visualized with profiles coloured by super-category. To capture non-linear structure, UMAP was applied [27], generating 2-D embeddings preserving both local and global similarity profiles. For supervised feature selection, a Random Forest classifier [28] with 500 trees was trained on labelled profiles, and feature importance was computed via mean decrease in Gini impurity. The classifier was also used to generate predicted super-category labels for "Unknown" profiles. This step was performed independently of the downstream machine learning–based validation analyses, which employed multiple classifiers and cross-validation as described below.

**Robustness, Effect-Size, and Machine-Learning Validation Analyses**

To assess clustering robustness, we compared multiple unsupervised configurations across imputation strategies, distance metrics, linkage rules, and cluster numbers. Specifically, median, most-frequent, and k-nearest-neighbour imputation were evaluated in combination with Euclidean, cosine, and correlation distances and average or Ward linkage. For each candidate configuration, silhouette scores, davies-bouldin and calinski-harabasz indices were calculated as clustering quality metrics, and the best-performing configuration was retained for downstream validation. Bootstrap branch support was estimated for the dendrogram, and the fully imputed matrix was exported for statistical dendrogram validation.

To quantify pairwise disease similarity more rigorously, we performed correlation analyses with uncertainty estimation. Pearson correlations were accompanied by Fisher-transformed confidence intervals and effective sample sizes, and sensitivity analyses were also conducted using Spearman and Kendall rank-based statistics. Feature-level contrasts were evaluated by comparing hematological disease profiles against non-hematological profiles, and each disease system against the remainder of the dataset. Standardized effect sizes were calculated using Hedges' g, with statistical significance assessed using Welch's t-test where relevant, after a Shapiro-Wilks test and Levene test assessed normality and homogeneity of variance. Multiple testing across features within each contrast was controlled using the Benjamini–Hochberg procedure (See Supplementary Table S2).

**Post-Hoc Machine Learning Classifiers: Multiclass Validation**

For post-hoc machine learning (ML) validation and Algorithmic Information Dynamics (AID) analyses, a complementary feature matrix was derived from a structured clinical knowledge resource used to obtain our disease-feature matrix, which encoded the 103 disease conditions across 10 organ systems (haematological, nephrology, endocrine/metabolic, cardiovascular, infective, gastrointestinal/hepatobiliary, rheumatological/vasculitis, neurological, respiratory, and oncology). Each condition was represented by binary and categorical features reflecting the utility of the full blood count (FBC) under three sampling paradigms: longitudinal sampling without a predefined signature, single sampling with a defined signature, and longitudinal sampling with a defined signature, across four diagnostic contexts: screening, possible diagnosis, diagnosis with additional tests, and disease monitoring. Analyte names were normalized using regular expressions, and the resulting structured matrix was used as the input feature space for all downstream multiclass supervised classification and AID perturbation analyses, providing a richer, ontology-informed representation of disease categories than the analyte-only profile matrix used for primary clustering. The system-level labels derived from this resource served as ground-truth class labels for classifier training and evaluation. To evaluate generalizability of the learned structure, repeated stratified multiclass ML classification was performed using Random Forest (RF), XGBoost, support vector machine (SVM) with radial basis function (RBF) kernel, and multilayer perceptron (MLP) models. ML algorithms were implemented in Python using scikit-learn and gradient boosting classifier.

The ML analysis used repeated stratified 5-fold cross-validation with 5 repeats (25 folds total), and performance was summarized using balanced accuracy, macro-F1 score, and one-vs-rest macro-Area Under the Receiver Operating Characteristic (AUROC) as ML performance metrics. Hyperparameters were fine-tuned a priori: Random Forest used 800 trees with balanced class weights and minimum leaf size of 2; SVM-RBF used standardization followed by an RBF-kernel classifier with C = 1.0, gamma = scale, and balanced class weights; XGBoost used 500 trees, maximum depth 4, learning rate 0.03, subsample 0.9, colsample_bytree 0.9, and regularization $\lambda$ = 1.0; and the MLP used two hidden layers (128, 64), ReLU activation, Adam optimization, early stopping, and a maximum of 600 iterations. In addition, leave-group-out sensitivity analyses were performed by removing hematology-related feature groups and reassessing silhouette scores, and explainable feature robustness/importance analysis was further evaluated using Gini importance, permutation importance, and SHAP (SHapley Additive exPlanations).

**Sensitivity Analysis: Imputation, Noise, and Embedding Stability**

To further evaluate the statistical robustness of the feature space and clustering structure, we performed a multi-stage sensitivity analysis incorporating imputation strategy, stochastic embedding variability, and noise perturbation analysis.

Missing data were handled using three imputation strategies implemented via scikit-learn: k-nearest neighbors (KNN), median imputation, and most-frequent imputation. Following imputation, all features were standardized and subjected to PCA using NumPy and SciPy-backed routines to assess global variance structure and separability across clinical categories. To characterize nonlinear manifold stability, UMAP was performed using the umap-learn Python library under multiple random initializations (seeds from 0 to 20). Default parameters were retained (n_neighbors and min_dist consistent across runs), isolating stochastic initialization as the primary source of variation (See Supplementary Information).

Noise robustness was assessed by injecting Gaussian noise into the feature matrix with $\sigma$ (sigma) defined relative to feature standard deviation (i.e., sigma scaling applied proportionally across features). For each noise level, embeddings and variance structure were recomputed, and stability was quantified via preservation of cluster topology and variance explained. All visual analyses were conducted in Python using pandas, matplotlib, and seaborn for visualization.

**Algorithmic Information Dynamics**

Algorithmic Information Dynamics (AID) was implemented as an in silico causal inference post-hoc network analysis using the pyBDM package to estimate changes in graph algorithmic complexity under feature perturbations. Features were binarized across multiple thresholds and evaluated using block decomposition method (BDM), Shannon entropy, and compression metrics (gzip, bz2, lzma, zlib) implemented via Python's standard libraries. Perturbation-based sensitivity was quantified as the change in information content (bits) following feature-wise perturbations, and features were ranked based on mean absolute shifts in information content across binarization thresholds and complexity measures.

## Results

Hierarchical clustering of the 103 disease profiles generated a dendrogram summarizing pairwise similarities across the dataset. Two complementary visualizations were produced: a full-scale tree covering the complete distance range (0.0–0.6; Figure 1) and a zoomed tree restricted to the high-similarity region (0.0–0.07; Figure 2). Together, these representations allow inspection of both global and fine-grained structure within the disease space. At the full scale, the tree demonstrates the broad heterogeneity of disease signatures, with branches spanning a wide spectrum of distances. Certain profiles appear as long, isolated branches, reflecting weak correlations with the majority of other conditions, in contrast to several groups of profiles that form tightly linked subtrees with minimal branch lengths, indicating high correlation and internal consistency. The zoomed view highlights the fine resolution of clustering at short distances, where closely related profiles merge into compact groups, emphasizing the presence of disease pairs or small clusters with very high similarity (ρ > 0.93), which are not as apparent in the full-scale visualization. This level of detail reveals that while some categories exhibit highly cohesive clustering, many profiles are distributed more diffusely across the tree, pointing to variable degrees of internal relatedness.

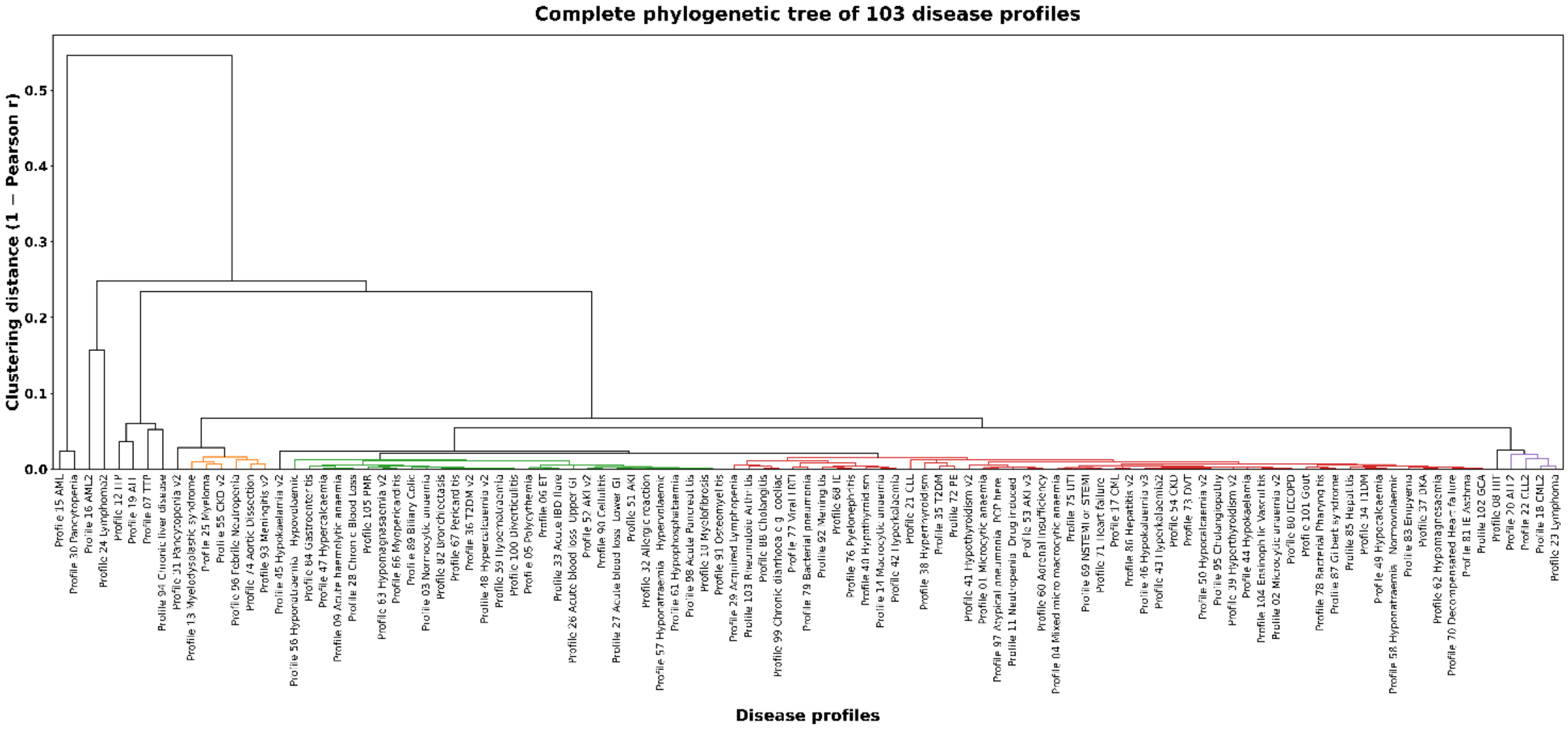

Figure 1. Complete phylogenetic tree of 103 disease profiles constructed using Pearson correlation distance ($1 - \rho ij$) and UPGMA hierarchical clustering (average linkage). The y-axis represents clustering distance.

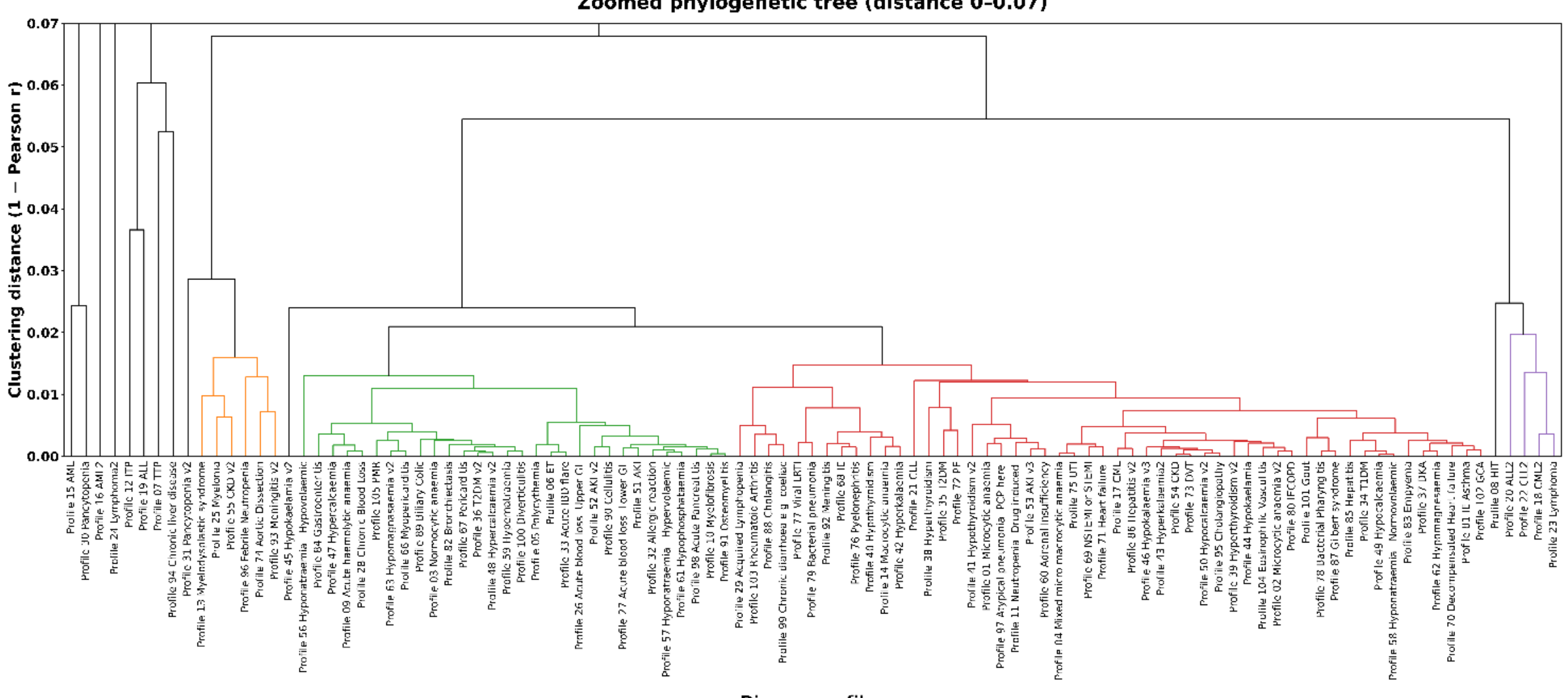

Figure 2. Phylogenetic tree of 103 disease profiles- Zoomed phylogenetic tree (distance 0–0.07), highlighting fine-scale relationships among closely related disease profiles.

## Cluster Analysis Outcomes

Cutting the hierarchical tree at a distance threshold of 0.02 produced 16 discrete clusters encompassing all 103 disease profiles. The distribution of cluster sizes was highly uneven, ranging from large, heterogeneous groupings to single-profile clusters. Most profiles (n = 80, 77.7%) aggregated within a single cluster (Cluster 9), which encompassed a wide spectrum of conditions spanning hematopoietic, metabolic, endocrine, cardiovascular, respiratory, gastrointestinal, and immune-related categories. This concentration indicates a 3.5-fold overrepresentation relative to all other clusters combined. The breadth of Cluster 9 indicates a high level of cross-system similarity among many disease signatures, despite their diverse clinical classifications. This heterogeneity motivated further analysis of its composition and potential shared mechanisms.

Outside of Cluster 9, most clusters were small. Several comprised a single disease profile, such as Cluster 1 (acute myeloid leukemia, AML), Cluster 2 (pancytopenia), Cluster 7 (myelodysplastic syndrome), and Cluster 16 (chronic liver disease). These singleton clusters reflect profiles that were relatively distinct from the remainder of the dataset at the chosen threshold. A few clusters contained two to six profiles, often representing conditions with moderate similarity. For example, Cluster 5 grouped viral lower respiratory tract infection with bacterial pneumonia, suggesting partial convergence among respiratory infections. Cluster 6 combined six conditions spanning hematopoietic, renal, cardiovascular, and infectious categories, underscoring cross-system associations. Cluster 10 grouped together chronic hematologic malignancies (CML2, CLL2, lymphoma), which aligns with known clinical relatedness.

Thus, cluster partitioning revealed a spectrum of outcomes: a single large, heterogeneous cluster (Cluster 9), several small clusters with modestly related diseases, and multiple singleton clusters representing distinct outliers. These results confirm that while certain categories, such as hematopoietic malignancies, can form coherent groups, the majority of disease profiles cluster into a broad, mixed set. This variability sets the stage for subsequent enrichment analyses to evaluate whether the observed clusters correspond to biologically meaningful groupings. A complete list of disease profiles within each of the 16 clusters is provided in

Appendix Table S1.

**Enrichment Analysis Findings**
To investigate whether large heterogeneous clusters captured biologically meaningful relationships, enrichment analyses were performed on Cluster 9, which contained 80 disease profiles spanning hematopoietic, metabolic, endocrine, cardiovascular, respiratory, gastrointestinal, and immune-related conditions. Gene–disease associations for these profiles were retrieved from DisGeNET, and the consolidated gene set was subjected to KEGG and Reactome over-representation analysis. Results are presented in Figures 2–8.

**KEGG Pathway Analysis**
To evaluate shared mechanisms within the largest heterogeneous cluster (Cluster 9), KEGG over-representation analysis was performed and the top results summarized in a dotplot (Figure 2). In this figure, p-values indicate raw significance from the hypergeometric test, whereas adjusted p-values (hereafter referred to as q-values) represent values corrected for multiple testing using the Benjamini–Hochberg false discovery rate procedure. The dot colour encodes q-values (more significant results appear toward the red end of the scale), dot size reflects the number of genes mapped to each pathway, and the x-axis shows the gene ratio.
The most highly enriched KEGG pathways included Cytokine–cytokine receptor interaction, Lipid and atherosclerosis, AGE–RAGE signaling pathway in diabetic complications, Rheumatoid arthritis, Th17 cell differentiation, Leishmaniasis, Inflammatory bowel disease, and Malaria. As indicated by the colour legend in Figure 2, q-values for these top pathways span from approximately $4.3 \times 10^{-15}$ (e.g., Th17 cell differentiation) down to approximately $9.4 \times 10^{-23}$ (e.g., Cytokine–cytokine receptor interaction). This gradient confirms that all reported pathways are highly significant, with several reaching extremely small q-values. Such extreme q-values are consistent with exceptionally large enrichment effects, indicating strong biological convergence across immune and inflammatory signaling pathways. Collectively, the KEGG results capture both broad immunological signaling themes and clinically recognizable disease processes, providing a coherent mechanistic context for the cross-system clustering observed in Cluster 9.

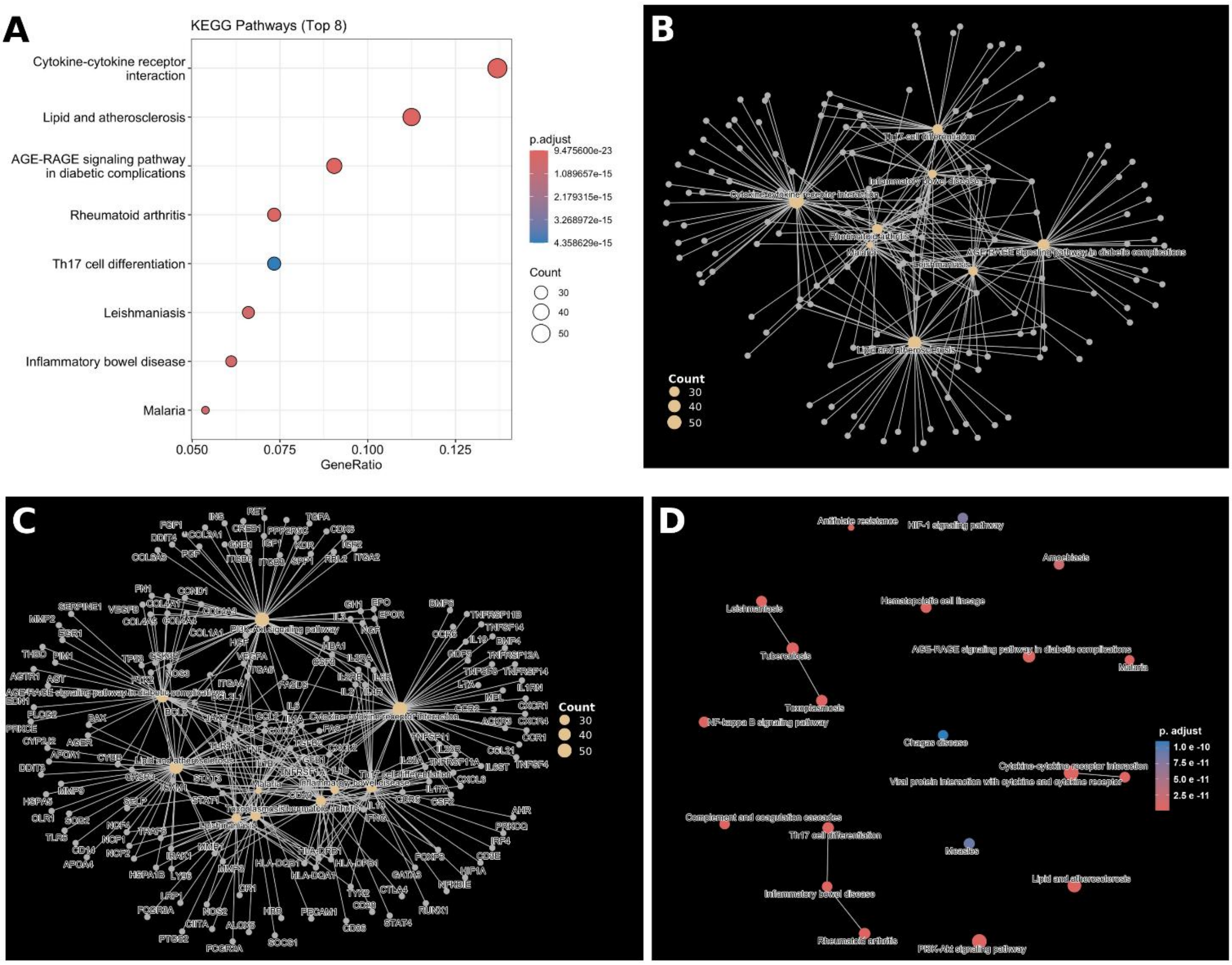

Figure 3. A) KEGG enrichment dotplot of Cluster 9 diseases. The top 8 pathways include Cytokine–cytokine receptor interaction, Lipid and atherosclerosis, AGE–RAGE signaling in diabetic complications, Rheumatoid arthritis, Th17 cell differentiation, Leishmaniasis, Inflammatory bowel disease, and Malaria. B) KEGG cnetplot of the top 8 enriched pathways, illustrating gene overlapping across immune, inflammatory, and infection-related processes. C) KEGG cnetplot (stringent threshold $q < 10^{-6}$), showing the gene–pathway network and hub genes connecting multiple pathways. D) KEGG emapplot (cutoff = 0.4), showing functional modules of interconnected pathways, including infection-related, autoimmune, and cytokine signaling clusters.

### KEGG cnetplot

Two complementary KEGG cnetplots were generated: (i) a network restricted to the top 8 enriched pathways (Figure 3A), and (ii) a broader network containing all pathways meeting a stringent significance threshold ($q < 1 \times 10^{-6}$; Figure 3B). In both panels, circular nodes represent pathways, smaller nodes represent genes, node size (for pathways) reflects the number of mapped genes, node colour encodes adjusted p-values (q-values), and edges indicate gene–pathway membership. Because no expression fold changes were supplied, gene node colours do not indicate differential regulation. In the top-8 cnetplot (Figure 3A), the eight pathways, such as Cytokine–cytokine receptor interaction, Lipid and atherosclerosis, AGE–RAGE signaling pathway in diabetic complications, Rheumatoid arthritis, Th17 cell differentiation, Leishmaniasis, Inflammatory bowel disease, and Malaria, form a compact, highly interconnected network. Multiple genes are shared across several of these terms, visually apparent as gene nodes linked to more than one pathway, which creates a bridge between the immune/inflammatory module (e.g., cytokine signaling, Th17 differentiation, rheumatoid arthritis, inflammatory bowel disease) and infection-related terms (e.g., leishmaniasis, malaria). The metabolic–vascular pathways (Lipid and atherosclerosis, AGE–RAGE in diabetic complications) are not isolated; instead, they connect into the same mesh via shared inflammatory mediators, consistent with the broader immune–inflammation theme observed in the dotplot (see Figure 2). The $q < 1 \times 10^{-6}$ cnetplot (Figure 3B) expands this view, retaining the relationships seen in the top-8 network while introducing additional significant pathways. The resulting graph is larger and denser, revealing sub-modules in which immune signaling and infection pathways cluster together through overlapping gene sets. This broader network confirms that the core structure observed in the top-8 panel is not an artifact of pathway selection; instead, it persists and becomes more granular when all highly significant terms are included. Practically, the top-8 panel offers a readable summary of the core mechanism, whereas the q-filtered panel provides a comprehensive map of pathway–gene connectivity at stringent significance.

**KEGG emapplot**

The emapplot (Figure 3C) provides a network physiology-based view of complex relationships among significantly enriched KEGG pathways, with edges indicating gene overlap greater than 40%. Node colour reflects adjusted p-values (q-values), node size corresponds to the number of mapped genes, and edges denote pathways sharing a substantial proportion of genes. Several functionally coherent modules emerged. First, an infection-related cluster was observed, linking Leishmaniasis, Tuberculosis, and Toxoplasmosis, all of which rely on overlapping immune defense genes. Second, an inflammation and autoimmunity module connected Th17 cell differentiation, Inflammatory bowel disease, and Rheumatoid arthritis, reflecting the central role of Th17-driven inflammatory processes in both systemic autoimmunity and gastrointestinal pathology. Third, a cytokine signaling module was evident, in which Cytokine–cytokine receptor interaction was directly linked with Viral protein interaction with cytokine and cytokine receptor, illustrating convergence between endogenous immune signaling and pathogen-driven modulation of cytokine networks. The majority of these nodes appeared in red or orange hues, consistent with extremely low q-values, indicating robust enrichment. Together, the emapplot highlights how the enriched KEGG pathways in Cluster 9 are not isolated entities but form interconnected modules centered on infection response, cytokine signaling, and chronic inflammatory processes.

**Reactome Pathway Analysis**

**Reactome dotplot**

The Reactome dotplot (Figure 3D) summarizes the top enriched pathways for Cluster 9. Node colour encodes adjusted p-values (q-values), node size reflects the number of mapped genes, and the x-axis denotes the gene ratio. The leading terms comprise an interleukin-centered immune axis—Signaling by interleukins, Interleukin-4 and Interleukin-13 signaling, and Interleukin-10 signaling—and a hemostasis/ECM axis—Platelet activation, signaling and aggregation, Platelet degranulation, Response to elevated platelet cytosolic $Ca^{2+}$, Extracellular matrix organization, and Degradation of the extracellular matrix. The predominance of warm hues (red/orange) indicates very low q-values across these pathways, consistent with robust enrichment. Together, the dotplot highlights two coherent Reactome themes in Cluster 9: cytokine/interleukin signaling and platelet/ECM biology.
Collectively, these cnetplots demonstrate that the Cluster 9 gene set is organized around a shared inflammatory and cytokine-mediated axis that links canonical immune pathways with infection-related and cardio-metabolic terms. The visible gene sharing across multiple pathways explains why heterogeneous clinical entities within Cluster 9 converge mechanistically at the level of KEGG annotations.

**Reactome cnetplot**

The Reactome cnetplot restricted to the top eight pathways (Figure 4A) visualizes gene–pathway membership as a bipartite network (pathways as larger circular nodes; genes as smaller nodes; edges indicate inclusion of a gene in a pathway; pathway-node colour encodes q-value; size reflects mapped-gene count). Three subgroups are apparent. First, an interleukin module is defined by Signaling by interleukins and Interleukin-4 and Interleukin-13 signaling, which lie in close proximity and share extensive gene overlap. Interleukin-10 signaling appears as a more isolated node: although positioned near the interleukin cluster, it is linked by fewer genes and thus forms a partially distinct subnode. Second, an extracellular matrix module is represented by Extracellular matrix organization and Degradation of the extracellular matrix, which cluster tightly together. Third, a platelet module encompasses Platelet activation, signaling and aggregation, Platelet degranulation, and Response to elevated platelet cytosolic $Ca^{2+}$, which are positioned in close proximity and indicate a coherent theme related to platelet function. This layout highlights that the Cluster 9 gene set maps onto discrete Reactome submodules, namely interleukin signaling, platelet activation, and matrix remodeling, with partial connectivity among them rather than a single unified hub.

The Reactome emapplots (Figure 4B and Figure 4C) present a pathway–pathway similarity network in which edges indicate substantial gene overlap (similarity cutoff $\approx$ 0.35), node colour encodes q-value, and node size reflects mapped-gene count. Three features are notable. First, a platelet triad forms a tight triangle among Platelet activation, signaling and aggregation, Platelet degranulation, and Response to elevated platelet cytosolic $Ca^{2+}$, all rendered in red, indicating very strong significance and high mutual overlap. Second, an interleukin cluster links Signaling by interleukins with Interleukin-4 and Interleukin-13 signaling, also in red, consistent with shared upstream/downstream mediators across interleukin pathways. Third, Degradation of the extracellular matrix (blue) connects to Extracellular matrix organization (red), indicating a functional pair with asymmetric significance: both participate in matrix remodeling, but the organization term is more strongly enriched.

**Dimensionality Reduction Insights**

**Principal Component Analysis (PCA)**

The PCA projection (Figure 5A) shows that PC1 explains 92.7% of the variance, with PC2 explaining 3.3%. Accordingly, the distribution of profiles is stretched predominantly along PC1, reflecting a single dominant axis of variation in the analyte matrix. Despite this, the two-component projection does not yield clean global separation between disease systems. Hematopoietic diseases (blue diamonds) exhibit a relatively narrow vertical spread along PC2, indicating partial convergence on secondary variance dimensions, while spanning an intermediate range across PC1 rather than aligning strictly with one side of the axis. By contrast, endocrine, metabolic, cardiovascular, respiratory, gastrointestinal, and other system categories remain broadly intermixed, clustering near the center of the projection with extensive overlap. These patterns suggest that although PCA captures dominant trends in the data, the variance structure is largely one-dimensional and insufficient on its own to resolve distinct system-level groupings.

**Uniform Manifold Approximation and Projection (UMAP)**

The UMAP embedding (Figure 5B) yields clearer structure than PCA, revealing three visually coherent zones along a curved manifold. On the rightmost island (UMAP1 ≈ 9–10), hematopoietic system diseases (blue diamonds) form a compact, internally consistent cluster with limited dispersion along UMAP2. A central zone (UMAP1 ≈ 6–8) contains a mixture of cardiovascular, urinary, gastrointestinal, endocrine, and other profiles, indicating cross-system proximity in blood-analyte space without a single dominant class. A leftward chain (UMAP1 ≈ –2 to 3) shows a graded progression that includes multiple non-hematopoietic super-categories together with several "Unknown" profiles, suggesting a continuous, non-linear spectrum rather than discrete partitions. Compared with PCA, UMAP thus enhances class coherence for hematopoietic diseases and reduces overlap among non-hematopoietic groups by capturing non-linear relationships.

**Random Forest Feature Importances**

A Random Forest classifier (500 trees; Gini impurity; feature importances normalized to sum to 1) was trained on profiles with known super-category labels. The top 15 features are summarized in the bar plot (Figure 11). Importances are distributed across a small set of hematological indices repeated over multiple timepoints, rather than concentrated in a single marker, indicating that several related analytes jointly drive discrimination. Segmented neutrophils (1000 cells/µL) appears four times among the top features (t5, t4, t1, t3), with t5 ranked first, indicating that absolute neutrophil counts carry strong class-discriminative signal. MCV (fL) appears at all measured timepoints (t1–t5) within the top 15, and RBC (million cells/µL) appears at t5, t1, and t2. Haemoglobin (g/dL) also enters the top 15 (t5). This recurrence across multiple timepoints indicates a dominant discriminative contribution of neutrophils relative to other analytes. Together, these features point to red-cell morphology and mass as major contributors to class separation. Platelet count (1000 cells/µL) is present at t4 and t5 among the highest-ranked features, complementing neutrophil and erythrocyte measures. Several late timepoints (t5) recur among the highest-ranked features, while earlier timepoints also appear, suggesting that the discriminative signal is stable across time rather than confined to a single measurement. The Random Forest highlights a hematology-centric signature: neutrophils, MCV, RBC, haemoglobin, and platelets repeatedly rank at the top across timepoints. This is concordant with earlier findings from clustering and non-linear embedding, in which hematopoietic disorders showed the most coherent grouping. We note that tree-based importances reflect model-specific attributions and can be diluted by collinearity among related features; hence, rankings should be interpreted as relative contributions rather than causal effects.

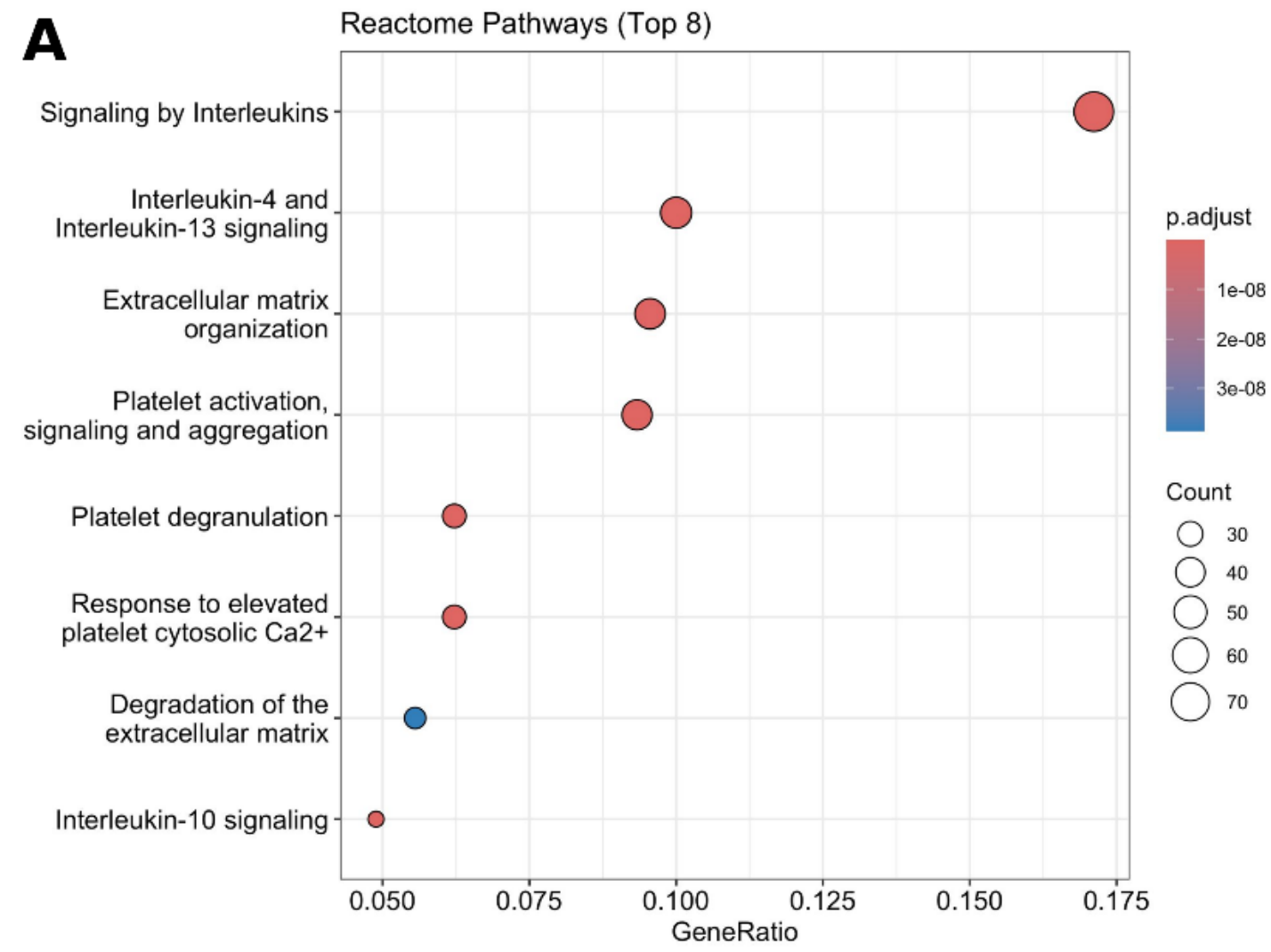

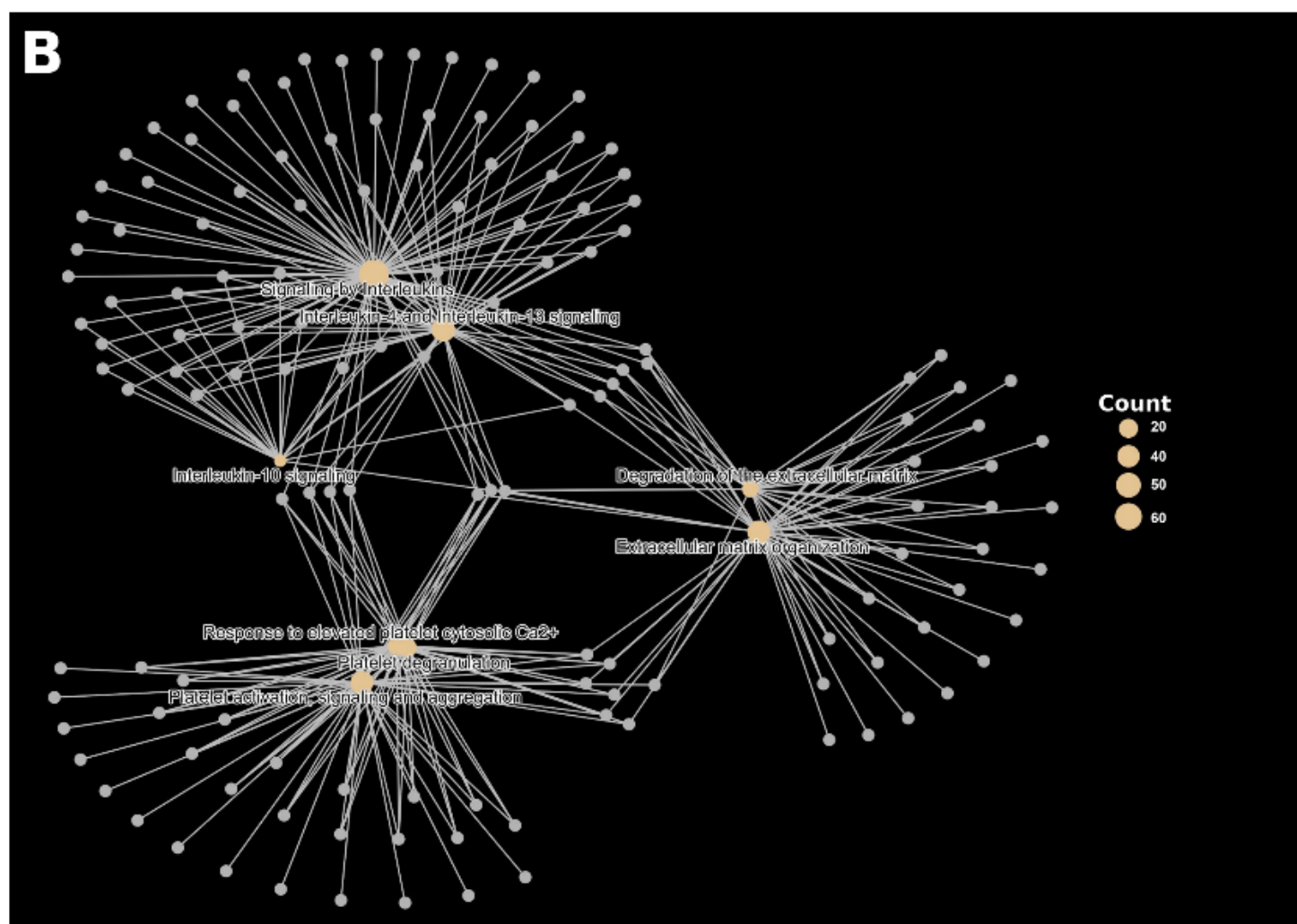

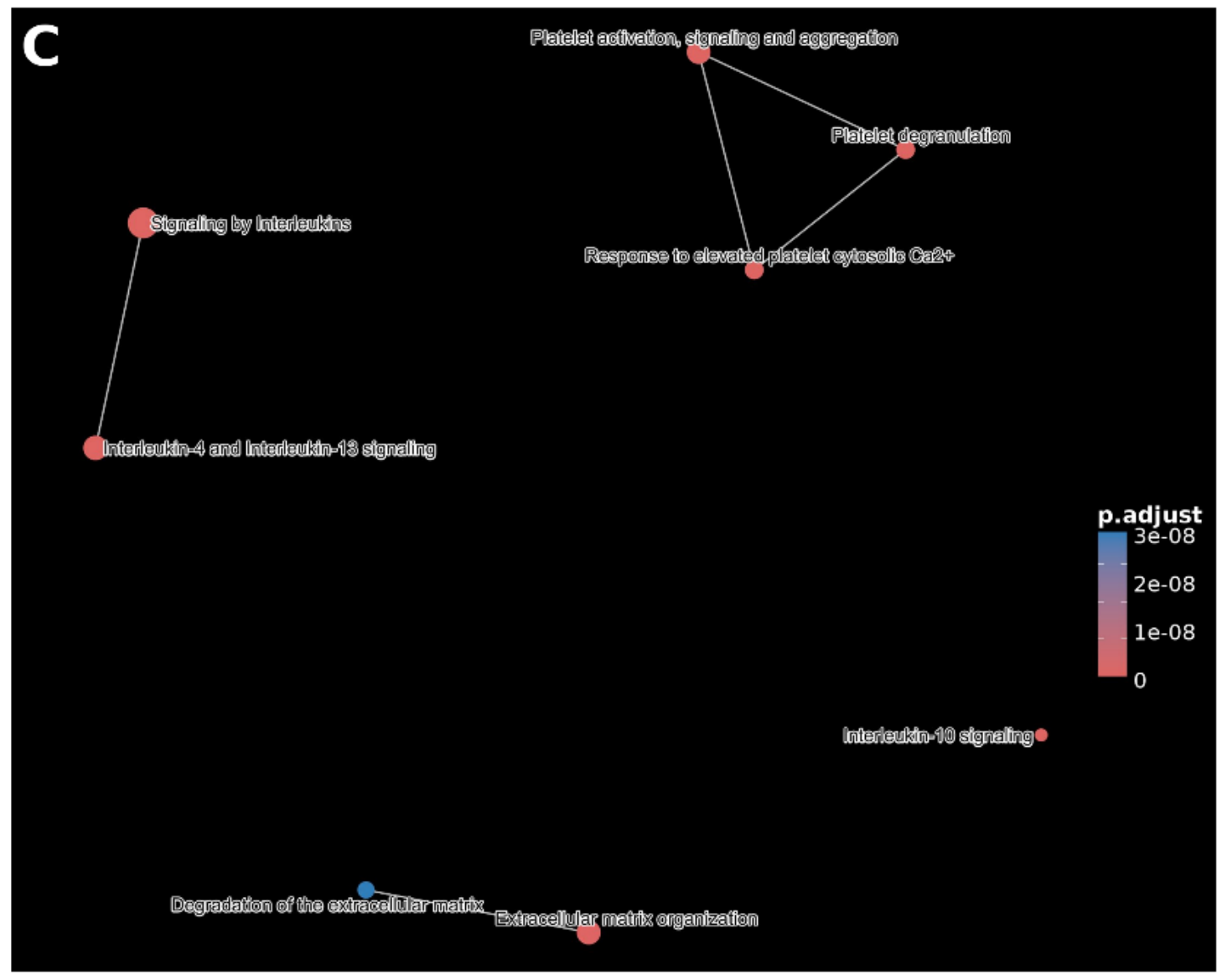

Figure 4. A) Reactome enrichment dotplot of Cluster 9 diseases. The top pathways include Signaling by interleukins, IL-4/IL-13 signaling, Extracellular matrix organization, Platelet activation/degranulation, and IL-10 signaling. B) Reactome cnetplot of the top 8 pathways, highlighting shared genes linking cytokine signaling, platelet function, and extracellular matrix remodeling. C) Reactome emapplot (cutoff = 0.35), showing modules

formed by platelet pathways, cytokine signaling, and extracellular matrix remodeling.

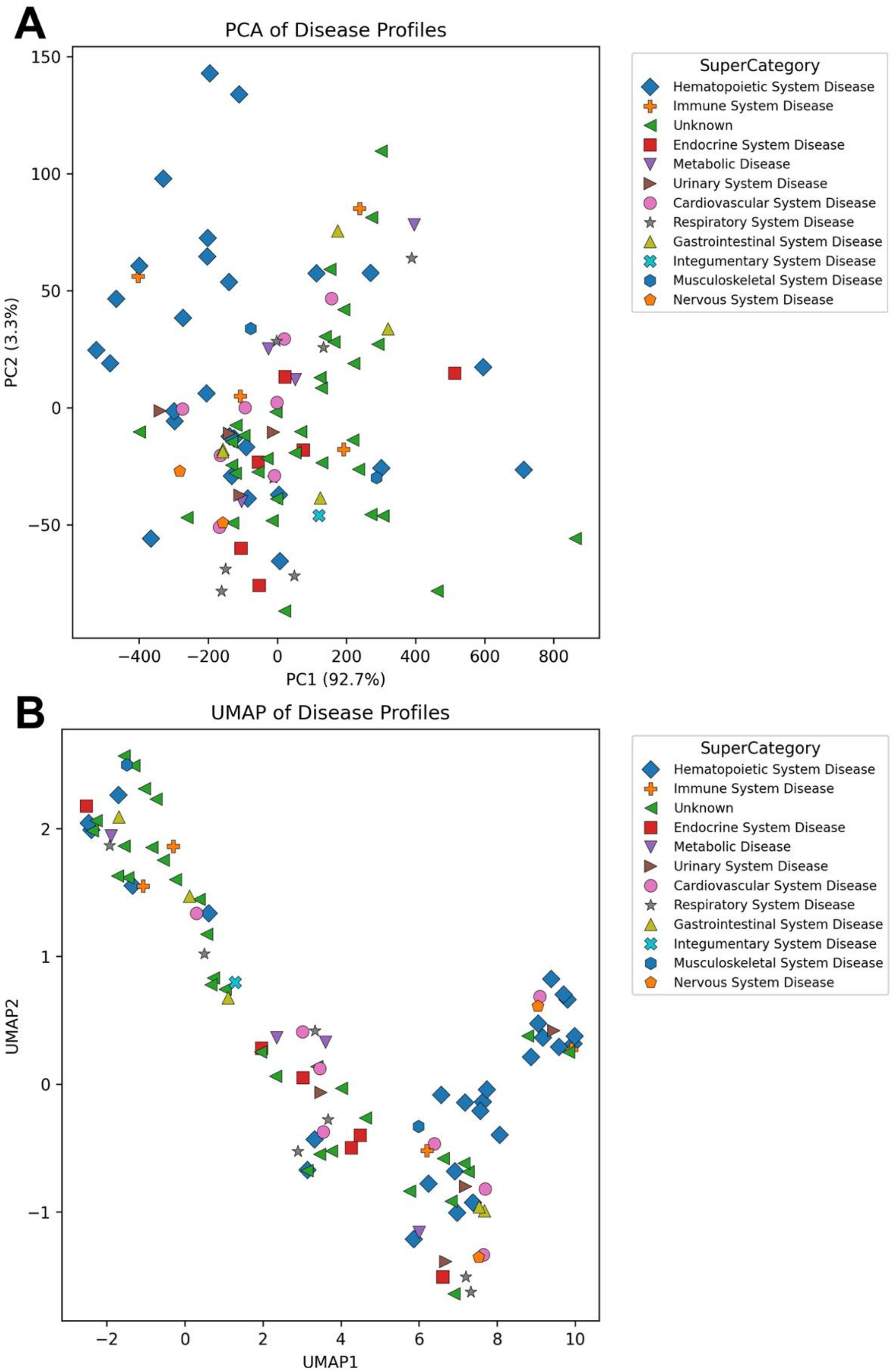

Figure 5. A) PCA of 103 disease profiles coloured by super-category. Hematopoietic diseases show vertical convergence along PC2, while non-hematopoietic systems remain dispersed. B) UMAP embedding of 103 disease profiles, showing a compact hematopoietic cluster separated from a continuous manifold of non-hematopoietic categories.

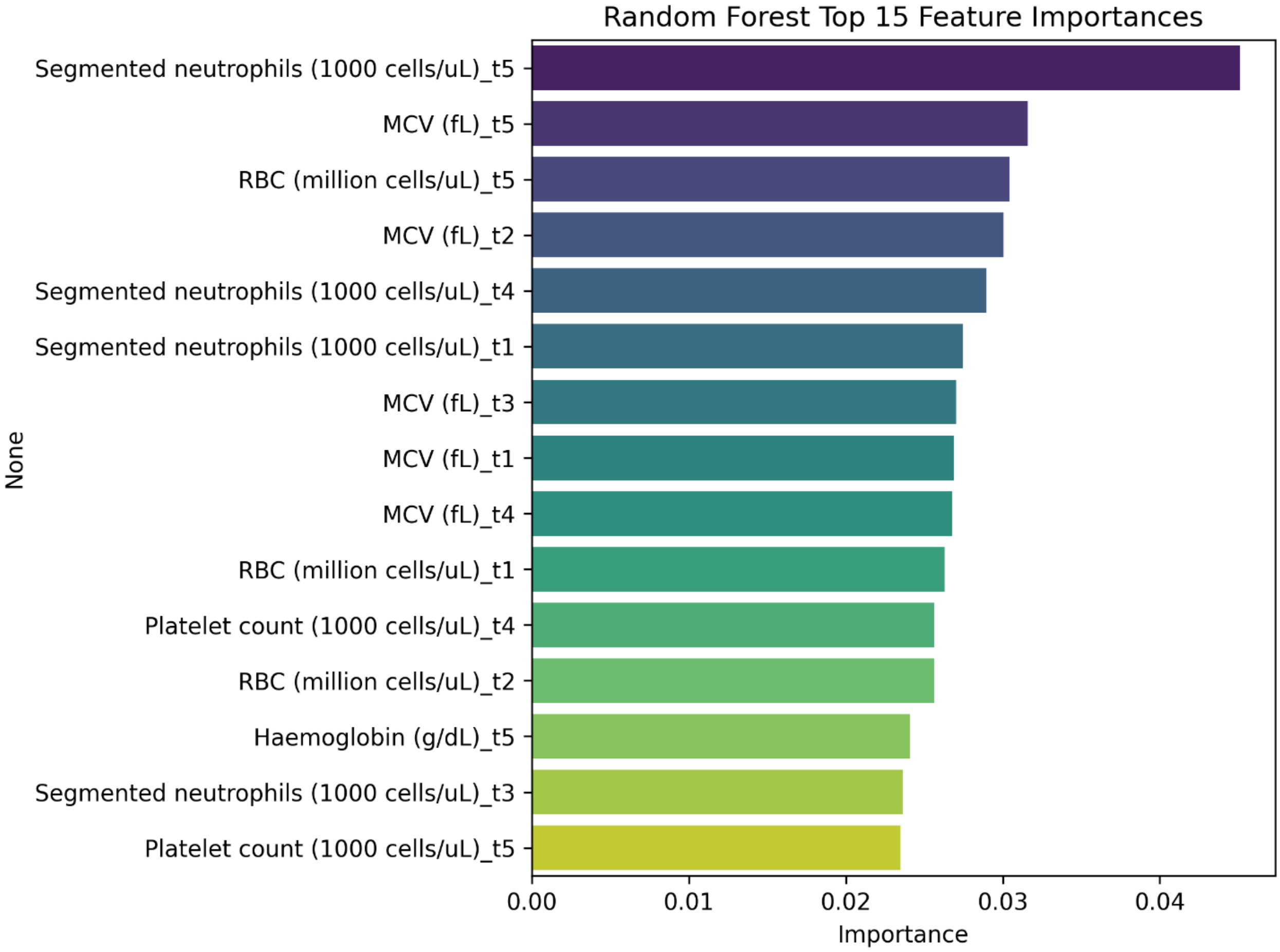

Figure 6: Random Forest analysis of feature importance. The top 15 features include segmented neutrophils, red blood cell indices (RBC, MCV, haemoglobin), and platelet counts, with hematological parameters dominating disease separation.

**Statistical Validation of Hierarchical Clustering (SVHC)**

To address concerns regarding the arbitrariness of dendrogram-based clustering, we applied Statistical Validation of Hierarchical Clustering (SVHC), an unsupervised clustering algorithm, which provides formal statistical testing of cluster significance through bootstrap resampling with FDR control on the intrinsic data structure. Importantly, clustering was performed on disease profiles ($n = 103$) represented as time-series hematologic feature vectors, rather than on analytes. Thus, the identified clusters reflect groups of disease states sharing modularity (hierarchical organization) among similar blood-based signatures as dynamical systems. Using 1000 bootstrap iterations ($\alpha = 0.05$), SVHC identified 25 statistically significant clusters ($p < 0.01$).

As shown in Figure 7, the validated clustering structure reveals both fine-grained subclusters and higher-order groupings, consistent with a multi-nested, multiscale organization of hematologic phenotypes. The correlation heatmap demonstrates pronounced diagonal block structures, indicating statistically robust intra-cluster similarity, alongside heterogeneous off-diagonal patterns reflecting inter-cluster relationships. These findings support the presence of modular organization within disease space, where subsets of conditions exhibit coherent hematologic profiles.

Clinically related disease categories, including inflammatory conditions, malignancies, and metabolic disorders, tended to co-localize within statistically validated modules, suggesting shared underlying biological signatures detectable from routine blood data. These results demonstrate that the observed clustering is not arbitrary, but reflects robust, statistically significant structure, reinforcing the presence of digital blood signatures as predictive biomarkers for modeling disease state-dynamics within a complex systems-level, data-driven framework and paving candidate targets for precision therapies in personalized medicine.

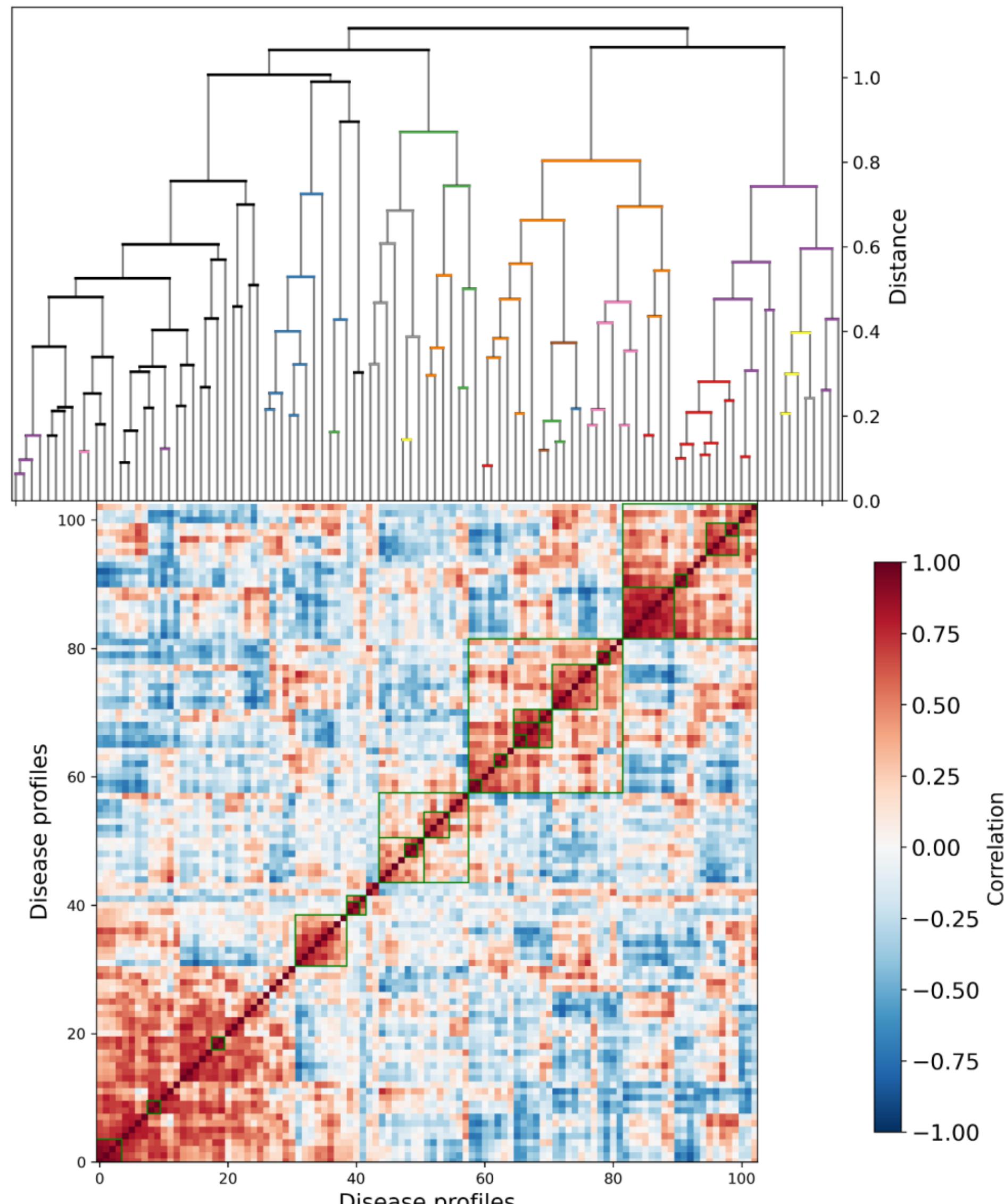

**Figure 7. Statistically validated hierarchical clustering of disease profiles using SVHC algorithm.**

Top: Hierarchical clustering dendrogram based on correlation distance between 103 disease profiles derived from time-series hematologic feature vectors. Bottom: Corresponding correlation matrix (Pearson), with red indicating positive correlation and blue negative correlation. Green boxes enriched along the diagonal denote statistically validated clusters identified by SVHC ($\alpha$ = 0.05, 1000 bootstrap samples, FDR-controlled). It shows 25 statistically validated clusters with distinct colour-coded branches at varying distance thresholds, and the heatmap reveals genuine heterogeneity including both strong positive (red) and negative (blue) correlations, indicating that the bootstrap resampling is discriminating real structure from noise. This modular organization supports the concept of digital blood signatures, suggesting that disease state-dynamics occupy structured regions in a biological state-space, akin to attractors in dynamical systems theory, enabling digital twin models to map, predict, and track patient-specific trajectories across these modules.

**Multiclass Classification and Feature Ablation Analysis**

To further quantify the robustness of disease groupings, we performed multiclass ML validation of disease separability and feature contribution analysis using simple classifiers. Across 25-fold cross-validation, ensemble methods demonstrated strong discriminative performance, with tree-based algorithms, namely Random Forest (balanced accuracy: 0.80 ± 0.13; AUROC: 0.97 ± 0.02) and XGBoost (balanced accuracy: 0.80 ± 0.13; AUROC: 0.97 ± 0.03) outperforming SVM (balanced accuracy: 0.71 ± 0.14; AUROC: 0.96 ± 0.03) and the basic neural network, MLP (balanced accuracy: 0.40 ± 0.12; AUROC: 0.78 ± 0.13). Class-wise ROC curves (Figure 8) showed near-perfect separability for hematological and infectious categories (AUC ≈ 1.00), with lower performance observed in nephrology and cardiovascular classes, reflecting greater heterogeneity and potential overlap in their hematologic profiles.

Feature ablation analysis further demonstrated the contribution of hematologic features to clustering structure. Removing hematology-related features increased the silhouette score from 0.22 to 0.27, indicating that while these features strongly drive ML classification performance, their removal enhances global cluster separation by reducing dominance of hematopoietic signals. This highlights a dual role of hematologic variables: they provide high discriminative power while simultaneously compressing broader inter-system variability.

Feature importance and robustness analyses further supported these findings, demonstrating consistent identification of key discriminative features across Gini importance, permutation importance, SHAP values, and bootstrap stability (Supplementary Table S5). The convergence of multiple ML-derived feature importance metrics indicates that classification performance is driven by stable, reproducible feature contributions rather than model-specific artifacts or chance effects. Together, with our statistical structure validation, these results reinforce the interpretation of disease phenotypes as modular regions within a shared hematologic state-space, with both high discriminability and biologically coherent variability across biosystems. These findings are consistent with a complex systems view in which disease states form separable yet interconnected attractor-like regions within a phenotypic landscape (i.e., routine blood feature state-space).

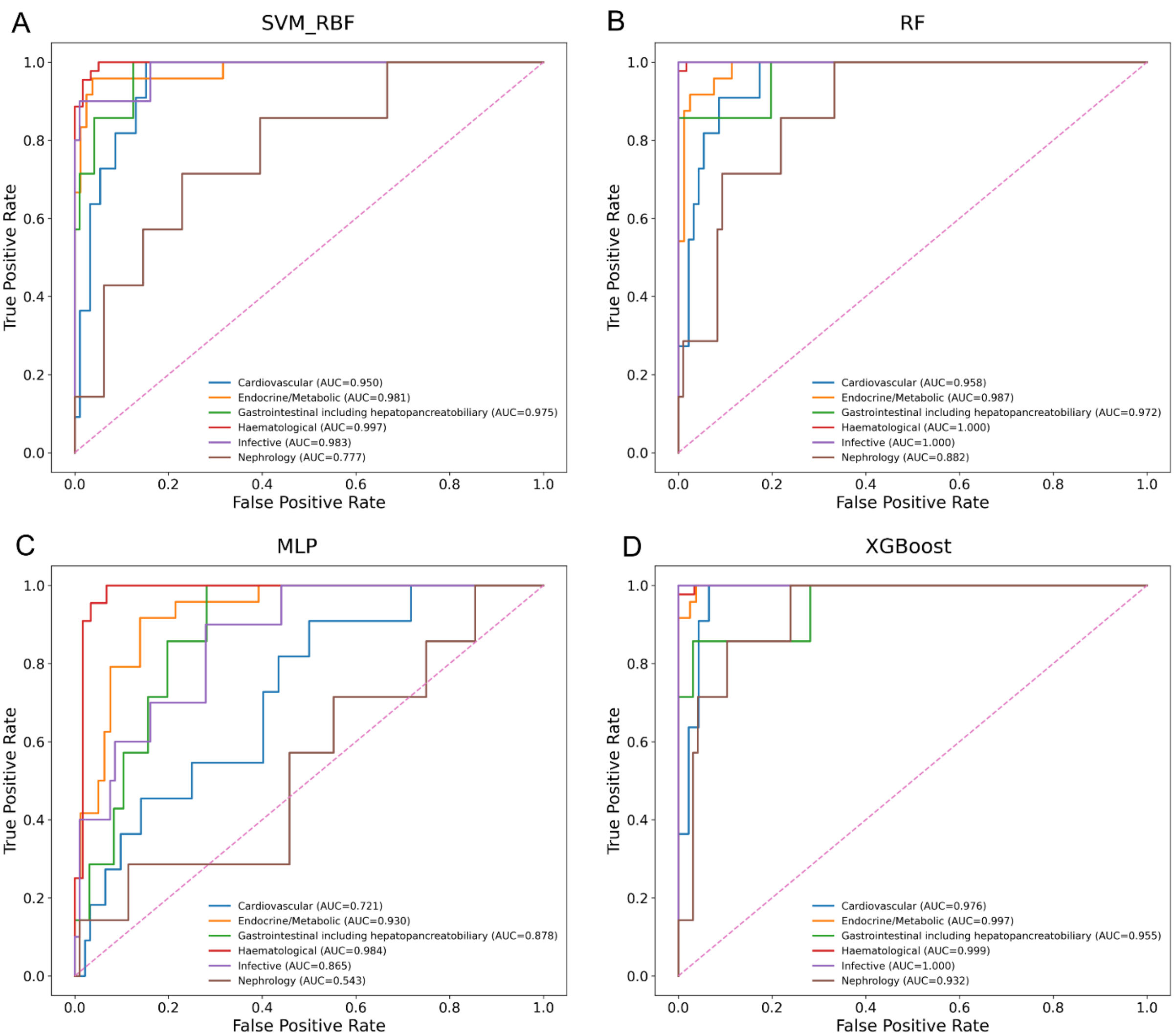

**Figure 8. Multiclass classification performance across disease systems.** Receiver operating characteristic (ROC) curves for A) SVM, B) RF, C) MLP and D) XGBoost. Ensemble methods (RF, XGBoost) achieved near-perfect discrimination for hematological and infectious diseases, while reduced performance in nephrology and cardiovascular classes. These results support the separability and robustness of disease clusters identified in unsupervised algorithms.

### Feature Space Sensitivity and Stability Analysis

Across all imputation strategies (KNN, median, most frequent), PCA demonstrated highly consistent variance structure, with PC1 explaining ~37–38% and PC2 ~12–13% of total variance. Visual inspection further confirmed that cluster organization and relative separation between disease groups were preserved across imputation methods, indicating minimal sensitivity to missing data handling (Figure 9 A-C).

UMAP embeddings also exhibited stable global organization across random seeds. While minor local rearrangements were observed as expected due to stochastic initialization of the UMAP algorithm, the global pattern/topology, including distinct clustering of haematological and endocrine/metabolic groups and transitional positioning of gastrointestinal and nephrology samples, remained highly conserved as seen by the colocalization of distinct clusters (colors). This indicates strong manifold stability and robustness of the learned feature space in nonlinear representation space (Figure 9 D-F).

Noise perturbation analysis further demonstrated resilience of the embedding structure. Increasing Gaussian noise resulted in gradual dispersion of clusters. However, core group separation persisted across moderate noise levels, suggesting that the signal captured by the feature space exceeds noise-induced variability. These findings confirm that both linear (PCA) and non-linear (UMAP) representations are robust to imputation choice, stochastic initialization, and moderate perturbations. Results in Table 2 and 3 complement Figure 9 by demonstrating that the observed pattern is robust not only to noise perturbation but also to preprocessing choices.

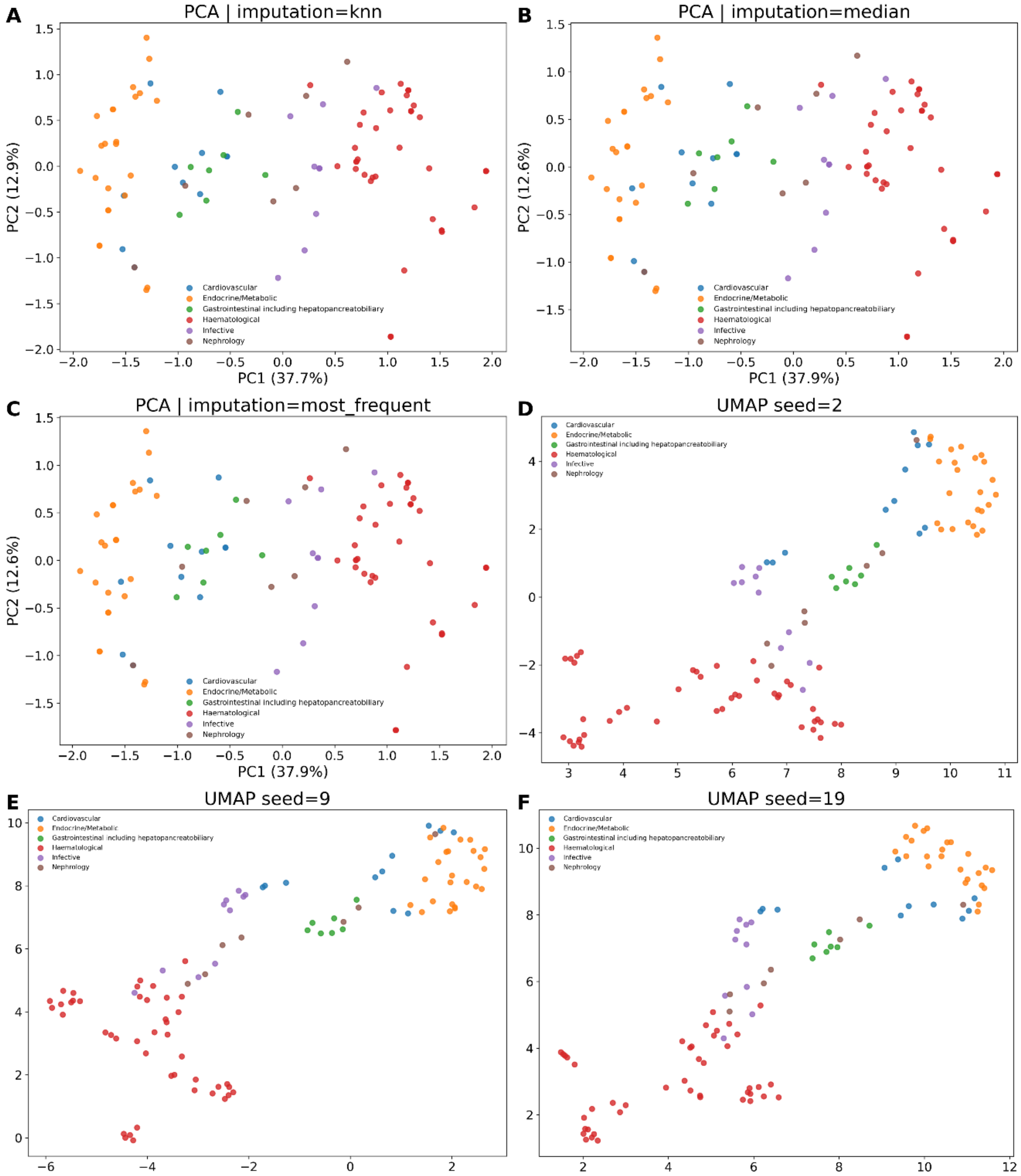

**Figure 9. Sensitivity analysis of feature space stability across imputation strategies and stochastic embedding initialization across learned global feature space.** (A–C) PCA under different imputation methods: (A) k-nearest neighbors (KNN), (B) median, and (C) most frequent imputation. Across all strategies, variance structure and inter-group relationships are preserved, indicating minimal sensitivity to missing data handling. (D–F) Uniform manifold

approximation and projection (UMAP) embeddings under different random seeds: (D) seed = 2, (E) seed = 9, and (F) seed = 19. Global topology and cluster organization remain stable across initializations.

| *Feature Space* | *σ* | *PC1 Variance* | *PC2 Variance* | *UMAP Procrustes Similarity* |
|---|---|---|---|---|
| *Time-series (55 features)* | 0.01 | 0.926566 | 0.033172 | 0.982791 |
| | 0.05 | 0.926545 | 0.03318 | 0.982791 |
| | 0.1 | 0.926542 | 0.033185 | 0.865897 |
| | 0.2 | 0.926524 | 0.03317 | 0.982791 |
| *Global feature space (ML/AID)* | 0.01 | 0.379231 | — | 0.721743 |
| | 0.05 | 0.374172 | — | 0.808008 |
| | 0.1 | 0.35432 | — | 0.949854 |
| | 0.2 | 0.305627 | — | 0.866766 |

**Table 2. Noise sensitivity analysis of time-series and learned global feature spaces**. Gaussian noise (σ = 0.01–0.2) was applied proportionally to standardized features in both the raw 55-feature longitudinal time-series matrix and the learned global feature space derived from machine learning/AID representations. Stability was assessed using variance explained by the first principal component (PC1; and PC2 for time-series) and structural preservation of UMAP embeddings via Procrustes similarity relative to the baseline (σ = 0).

The time-series feature space demonstrates extremely high variance preservation (PC1 ≈ 0.93) and near-complete embedding stability across all noise levels, indicating a highly dominant low-dimensional structure. In contrast, the learned global feature space exhibits lower initial variance concentration but maintains strong embedding stability across noise levels, particularly at moderate perturbations (σ = 0.1), reflecting a more distributed yet robust representation.

| **FEATURE SPACE** | **IMPUTATION** | **N FEATURES** | **PC1 VARIANCE** | **PC2 VARIANCE** | **PC1+PC2** | **PC1–PC5 (IF APPLICABLE)** |
|---|---|---|---|---|---|---|
| **TIME-SERIES (55 FEATURES)** | median | 55 | 0.926565 | 0.033172 | 0.959738 | 0.990834 |
| | most_frequent | 55 | 0.927421 | 0.03308 | 0.9605 | 0.990791 |
| | knn | 55 | 0.927543 | 0.033176 | 0.960719 | 0.990796 |
| **GLOBAL FEATURE SPACE (ML/AID)** | median | 25 | 0.378728 | 0.125739 | 0.504466 | — |
| | most_frequent | 25 | 0.378728 | 0.125739 | 0.504466 | — |
| | knn | 25 | 0.377301 | 0.128686 | 0.505987 | — |

**Table 3. Effect of imputation strategy on variance structure in time-series and learned global feature spaces.** Comparison of principal component variance across imputation strategies (median, most frequent, k-nearest neighbors) for both the raw 55-feature longitudinal time-series matrix and the learned global feature space derived from machine learning/AID representations. The proportion of variance explained by the first two principal components (PC1, PC2), along with cumulative variance, remains highly consistent across imputation methods in both spaces.

The time-series feature space is dominated by a single principal component (PC1 ≈ 0.93), with over 99% of variance captured by the first five components, indicating a highly compressible, low-dimensional structure in temporal ordering. This suggests that longitudinal trajectories are governed by a limited set of dominant transitions, consistent with a coarse-grained, two-state organization and attractor-like dynamics.

In contrast, the learned global feature space distributes variance more evenly (PC1 ≈ 0.38), reflecting a more complex and heterogeneous spatial organization of disease profiles. Together, these findings indicate that while temporal trajectories collapse onto a simplified, low-dimensional manifold, the global feature space captures a richer multiscale structure with greater separability between disease states. The consistency of these patterns across imputation strategies further indicates that the observed variance structure is preserved irrespective of missing data handling.

| *imputation* | *metric* | *k* | *silhouette* | *davies_bouldin* | *calinski_harabasz* |
|---|---|---|---|---|---|
| *median* | euclidean | 2 | 0.561723 | 0.490831 | 44.2551 |
| *most_frequent* | euclidean | 2 | 0.561157 | 0.491386 | 44.1427 |
| *knn* | euclidean | 2 | 0.560725 | 0.491493 | 44.0785 |
| *knn* | euclidean | 3 | 0.478343 | 0.629162 | 117.6468 |
| *most_frequent* | euclidean | 3 | 0.477878 | 0.629389 | 117.5446 |
| *median* | euclidean | 3 | 0.476089 | 0.632381 | 116.9803 |
| *knn* | euclidean | 5 | 0.46514 | 0.588128 | 126.4637 |
| *most_frequent* | euclidean | 5 | 0.464821 | 0.587927 | 126.3299 |
| *median* | euclidean | 5 | 0.463115 | 0.588951 | 125.4962 |
| *knn* | euclidean | 6 | 0.440943 | 0.592101 | 110.2459 |
| *most_frequent* | euclidean | 6 | 0.440582 | 0.592317 | 110.1301 |
| *median* | euclidean | 6 | 0.438695 | 0.592827 | 109.3612 |
| *most_frequent* | euclidean | 7 | 0.430705 | 0.556085 | 95.7941 |
| *knn* | euclidean | 7 | 0.42994 | 0.556675 | 95.8045 |
| *median* | euclidean | 7 | 0.428678 | 0.556771 | 95.0748 |
| *knn* | euclidean | 4 | 0.40758 | 0.582826 | 81.875 |
| *most_frequent* | euclidean | 4 | 0.407324 | 0.582982 | 81.8078 |
| *median* | euclidean | 4 | 0.405738 | 0.585198 | 81.4075 |
| *most_frequent* | correlation | 2 | 0.399579 | 0.453662 | 8.5238 |
| *median* | correlation | 2 | 0.39954 | 0.45438 | 8.5191 |
| *knn* | cosine | 2 | 0.360446 | 0.827791 | 11.7095 |
| *knn* | correlation | 2 | 0.360446 | 0.827791 | 11.7095 |
| *most_frequent* | cosine | 2 | 0.36016 | 0.835643 | 11.6656 |
| *median* | cosine | 2 | 0.360041 | 0.835066 | 11.6584 |
| *knn* | cosine | 3 | 0.336343 | 0.989001 | 14.2083 |
| *most_frequent* | cosine | 3 | 0.335951 | 1.008014 | 14.1796 |
| *median* | cosine | 3 | 0.335948 | 1.005906 | 14.1673 |
| *most_frequent* | correlation | 4 | 0.301842 | 0.732756 | 9.9511 |
| *most_frequent* | cosine | 4 | 0.301842 | 0.732756 | 9.9511 |
| *median* | cosine | 4 | 0.301556 | 0.733811 | 9.9403 |
| *median* | correlation | 4 | 0.301556 | 0.733811 | 9.9403 |
| *knn* | correlation | 4 | 0.300403 | 0.740142 | 9.9472 |
| *knn* | cosine | 4 | 0.300403 | 0.740142 | 9.9472 |
| *most_frequent* | correlation | 3 | 0.275342 | 0.802493 | 6.4519 |
| *median* | correlation | 3 | 0.274972 | 0.803911 | 6.4454 |
| *knn* | correlation | 3 | 0.273985 | 0.812369 | 6.4476 |
| *most_frequent* | correlation | 5 | 0.267612 | 0.516463 | 7.5788 |
| *most_frequent* | cosine | 5 | 0.267612 | 0.516463 | 7.5788 |
| *median* | correlation | 5 | 0.267305 | 0.517127 | 7.5709 |
| *median* | cosine | 5 | 0.267305 | 0.517127 | 7.5709 |
| *knn* | cosine | 5 | 0.266624 | 0.521944 | 7.5758 |
| *knn* | correlation | 5 | 0.266624 | 0.521944 | 7.5758 |
| *most_frequent* | correlation | 6 | 0.223157 | 0.601551 | 10.9965 |
| *median* | correlation | 6 | 0.222729 | 0.602327 | 10.9813 |
| *median* | cosine | 6 | 0.222729 | 0.602327 | 10.9813 |
| *knn* | correlation | 6 | 0.222109 | 0.606166 | 10.9934 |
| *knn* | cosine | 6 | 0.222109 | 0.606166 | 10.9934 |
| *most_frequent* | correlation | 7 | 0.21103 | 0.715018 | 9.0992 |
| *median* | correlation | 7 | 0.210604 | 0.715578 | 9.0866 |
| *median* | cosine | 7 | 0.210604 | 0.715578 | 9.0866 |
| *knn* | correlation | 7 | 0.209901 | 0.718961 | 9.0966 |
| *most_frequent* | cosine | 6 | 0.19186 | 0.613178 | 11.4809 |
| *most_frequent* | cosine | 7 | 0.175003 | 0.742828 | 9.4996 |
| *knn* | cosine | 7 | 0.142106 | 0.679437 | 11.3324 |

**Table 4. Clustering performance across imputation strategies and distance metrics (55-feature time-series space).** Clustering quality metrics computed on the full 55-feature longitudinal time-series matrix (103 disease profiles) across multiple imputation strategies (median, most frequent, k-nearest neighbors) and distance metrics (Euclidean, correlation, cosine). Performance was evaluated using silhouette score, Davies–Bouldin index, and Calinski–Harabasz index across cluster numbers (k = 2–7). Higher silhouette and Calinski–Harabasz scores and lower Davies–Bouldin values indicate improved clustering quality.

| Feature | BDM | BZ2 | Entropy | GZIP | LZMA | ZLIB |
|---|---|---|---|---|---|---|
| hematology_platelet | 99.65 | 5 | 0.08805 | 9.2 | 5.6 | 6.6 |
| neuro | 75.76 | 4.8 | 0.079013 | 6.8 | 4 | 6.6 |
| hematology_splenomegaly | 75.37 | 5 | 0.085973 | 9.2 | 4.8 | 8.2 |
| renal_urinary | 73.28 | 2.8 | 0.086201 | 6.8 | 4.8 | 7.6 |
| endocrine_metabolic | 69.33 | 6.8 | 0.08805 | 7.8 | 4 | 7 |
| hematology_leukemia_lymphoma | 66.66 | 5.8 | 0.08805 | 9.2 | 5.6 | 8.6 |
| gi_hepatobiliary | 58.83 | 3.2 | 0.067118 | 6 | 4 | 5 |
| monitor_longitudinal_with_signature | 58.33 | 7.6 | 0.059938 | 6.6 | 4 | 7.4 |
| monitor_single_with_signature | 57.04 | 4 | 0.068806 | 7.2 | 4 | 7.6 |
| fbc_dx_longitudinal_without_signature | 49.69 | 5.6 | 0.08805 | 7.8 | 5.6 | 8.4 |
| cardiovascular | 42.56 | 4.4 | 0.057641 | 6 | 4 | 3.8 |
| hematology_bleeding | 42.37 | 3.6 | 0.0778 | 7 | 5.6 | 8.8 |
| infection_inflammation | 40.14 | 3.4 | 0.062395 | 4.6 | 5.6 | 7.6 |
| respiratory | 37.92 | 3.4 | 0.067679 | 5.4 | 4 | 5.4 |
| monitor_longitudinal_without_signature | 37.43 | 7.4 | 0.056083 | 8.4 | 4 | 6.2 |
| screening_longitudinal_with_signature | 34.77 | 8.8 | 0.055019 | 6.8 | 5.6 | 6 |
| fbc_dx_longitudinal_with_signature | 32.89 | 5.6 | 0.078287 | 10.4 | 6.4 | 8 |
| hematology_anemia | 28.53 | 3.8 | 0.062537 | 4.4 | 3.2 | 4.4 |
| screening_single_with_signature | 28.30 | 6.6 | 0.052698 | 5.8 | 5.6 | 6.2 |
| fbcplus_dx_single_with_signature | 27.41 | 6.2 | 0.05191 | 6.6 | 4 | 7.8 |
| fbcplus_dx_longitudinal_without_signature | 27.13 | 7 | 0.053125 | 6 | 5.6 | 4.6 |
| screening_longitudinal_without_signature | 25.03 | 8.4 | 0.052168 | 4.8 | 4 | 5.8 |
| fbcplus_dx_longitudinal_with_signature | 23.46 | 4.4 | 0.054502 | 7 | 3.2 | 6.4 |
| fbc_dx_single_with_signature | 16.61 | 2.6 | 0.075449 | 9.4 | 6.4 | 8.8 |
| thrombosis | 0 | 8.6 | 0.086566 | 7.8 | 3.2 | 7.2 |

**Table 5. Top Global Features Identified by Algorithmic Information Dynamics (BDM and Compression Metrics Perturbation Analysis).**

To further characterize structural feature importance beyond statistical and machine learning–based metrics, we applied an AID framework incorporating BDM, Shannon entropy, and compression-based complexity measures (gzip, bz2, lzma, zlib). Features were systematically perturbed across multiple binarization thresholds, and the resulting change in algorithmic information content was quantified. BDM perturbation analysis identified a small subset of predictive features, particularly hematology_platelet, neuro, hematology_splenomegaly, renal_urinary, and endocrine_metabolic, as exhibiting the largest BDM shifts.

From a causal inference perspective, these predictive features can be interpreted as structurally influential components of the system, or disease signatures: perturbations to these signatures induce disproportionately large changes in the global information structure of the disease-feature matrix, or graph network. This suggests that they act as potential control or regulatory elements within the hematologic state-space, consistent with an attractor-like dynamics in which specific network features govern critical transitions and stability across disease phenotypes. These findings complement our supervised feature importance and clustering results, reinforcing that the observed disease structure is driven by a subset of biologically meaningful patterns rather than random signals (See Figure S3 in Supplementary Info).

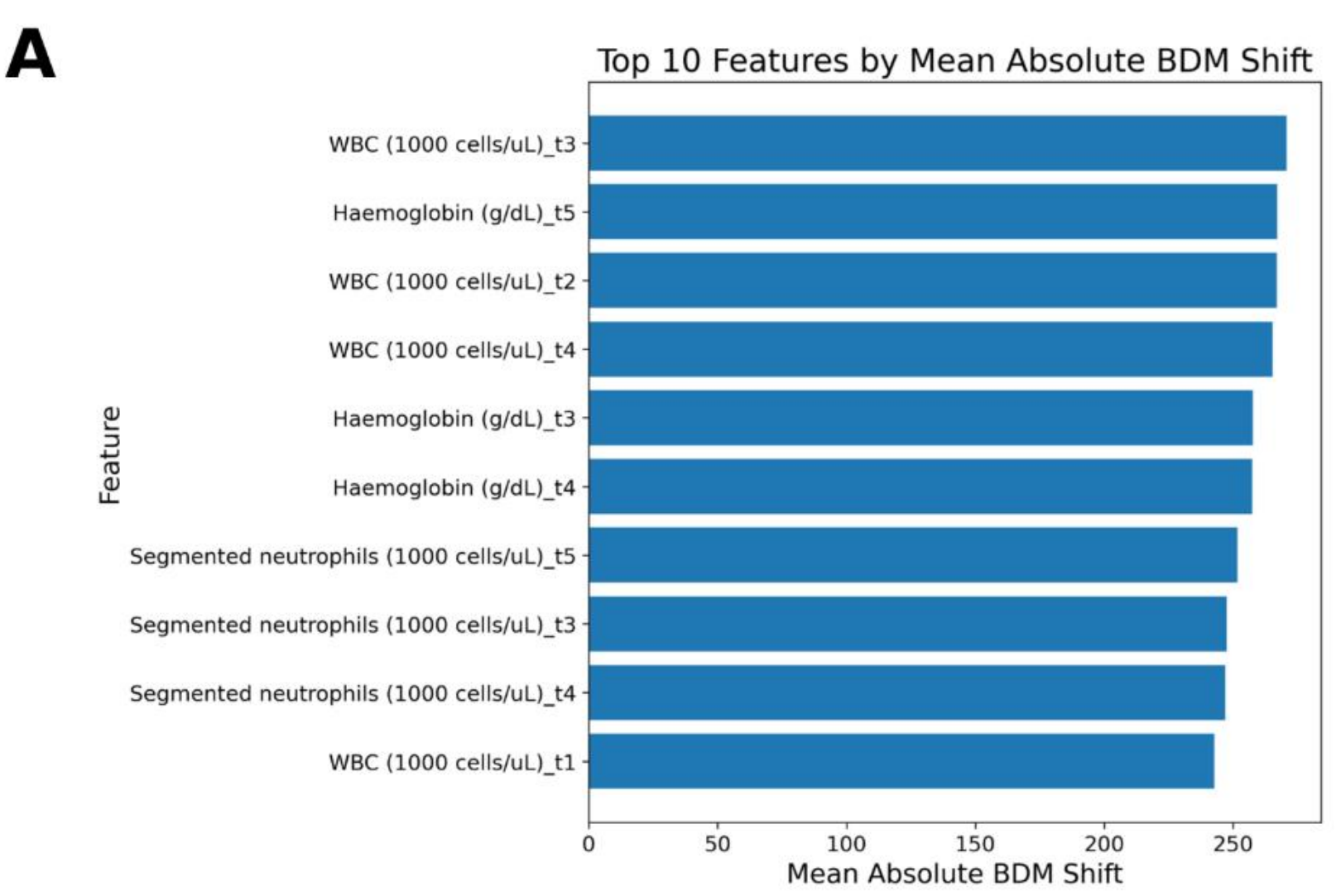

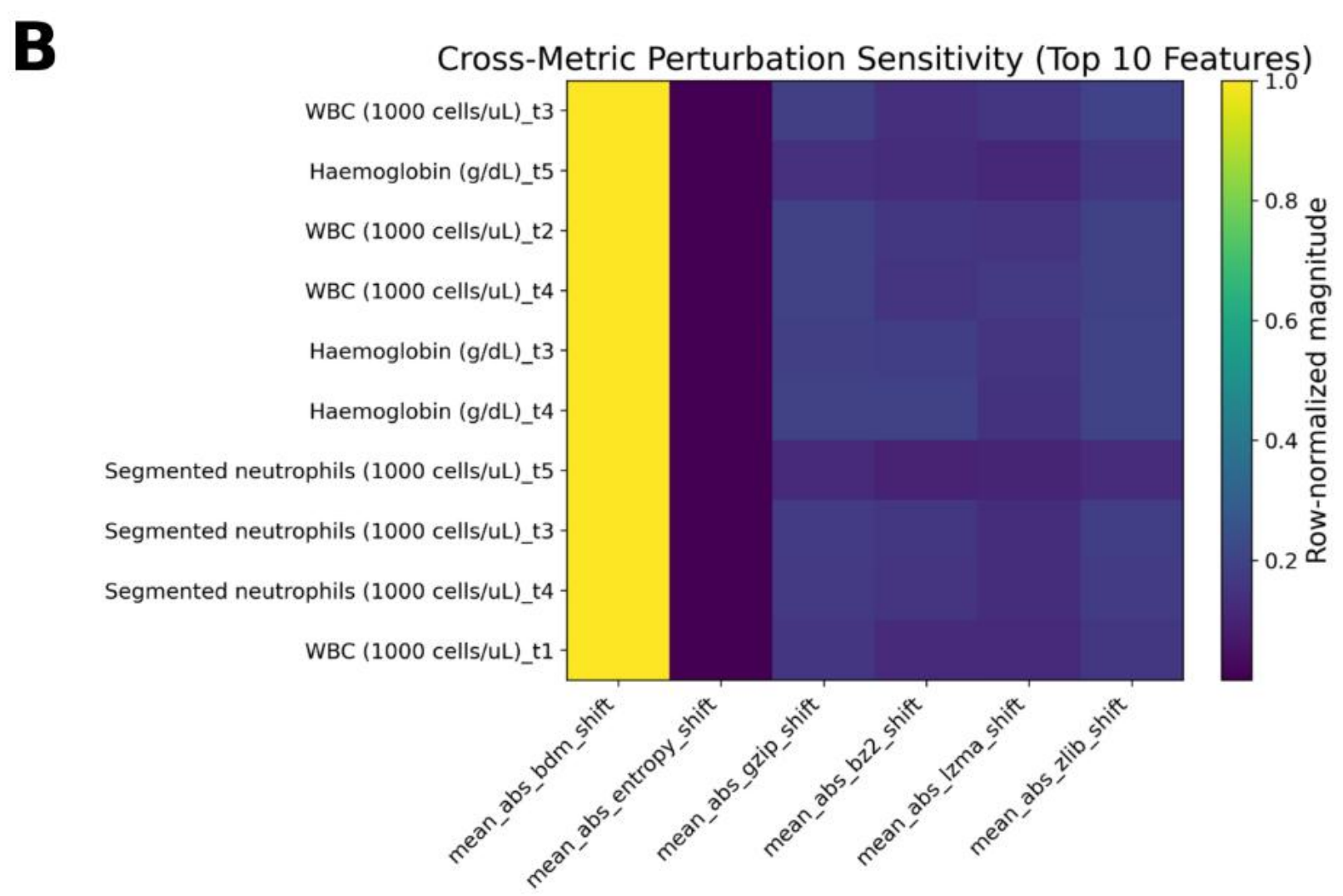

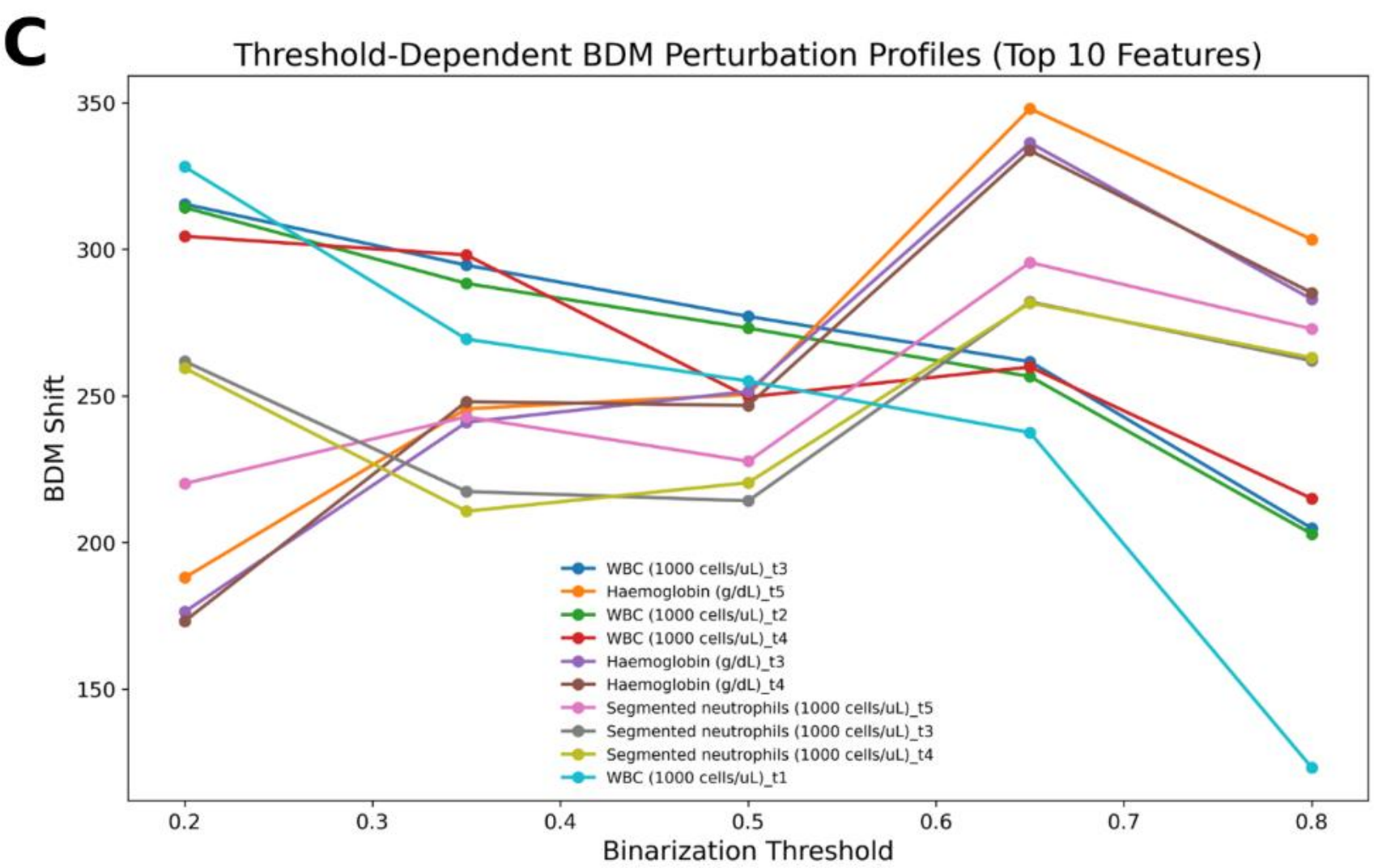

**Figure 10. AID-Based Perturbation Analysis Identifies Dominant Control Features in the**

**Time-Series Disease Feature Space. (A)** Top 10 features ranked by mean absolute BDM shift, highlighting the most structurally influential variables in the global information landscape. **(B)** Cross-metric perturbation sensitivity heatmap (row-normalized) comparing BDM, entropy, and compression-based complexity measures (gzip, bz2, lzma, zlib) for the top features. **(C)** Threshold-dependent BDM perturbation profiles across binarization thresholds (0.2–0.8), demonstrating consistency and robustness of the top 10 BDM shift features.

Algorithmic Information Dynamics (AID) perturbation analysis reveals a small subset of hematologic features, primarily WBC, haemoglobin, and segmented neutrophils across multiple timepoints, that exert disproportionate causal influence on the global information structure. BDM-based complexity shifts consistently dominate over entropy and compression-based metrics, indicating superior sensitivity to structural perturbations (Figure 10 A-C). The recurrence of these hematological features across thresholds and timepoints among the diverse disease groups suggests temporally stable, complexity signatures robust to noise perturbations or randomness. Collectively, these results support an attractor-like self-organization of the routine blood data-derived disease feature space, where a limited set of predictive features governs transitions within a structured manifold (See Figure S4 in Supplementary).

## Discussion

### Hematopoietic clustering

A central and consistent observation across all analytical approaches was the robust clustering of hematopoietic diseases. In the phylogenetic tree, disorders such as acute and chronic leukemias, lymphomas, myelodysplastic syndromes, and anemias either formed discrete compact subclusters or appeared as isolated but internally coherent branches. In correlation-based clustering, these profiles repeatedly grouped together, in stark contrast to the dispersion of most endocrine, metabolic, or cardiovascular diseases within a large heterogeneous cluster. Dimensionality reduction reinforced this distinction: in PCA, hematopoietic conditions exhibited a relatively narrow vertical spread along PC2, indicating convergence on secondary variance dimensions, while spanning an intermediate range across PC1 without forming complete separation from other systems. In UMAP, by contrast, they formed a particularly tight and distinct cluster that was clearly segregated from the broader manifold of non-hematopoietic categories. Random Forest classification further underscored this pattern by identifying neutrophil counts, red blood cell indices (MCV, RBC), and platelet counts as the top discriminative features, pointing to classical hematological markers as the most informative drivers of disease separation.

The consistency of hematopoietic clustering across correlation metrics, enrichment, dimensionality reduction, and feature selection methods, as well as the algorithmic information dynamics-driven perturbation analyses, underscores the stability of blood-based phenotyping for these conditions. It also highlights a contrast with other complex systems: whereas hematopoietic disorders form coherent entities in analyte pattern space, endocrine, metabolic, and cardiovascular profiles appear more diffuse, reflecting greater heterogeneity and overlap. This divergence also suggests that blood-based disease similarity measures may be most directly applicable to disorders of the hematopoietic system, and that expanding their interpretability to other systems will require integrating additional modalities ranging from subjective data enrichment (signs and symptoms) to emerging multi-omics frameworks, including genetic (e.g., spatial transcriptomics), or tissue-specific precision biomarkers.

However, it should also be noted that the prominence of hematopoietic clustering observed across methods is, in part, expected given that the input data consists of blood-derived analytes, which are encompassed within hematologic processes. Yet, this structure may also be modality-dependent: blood-based features preferentially capture systemic and hematologic perturbations, while other disease classes are reflected indirectly through secondary effects, and whose biomarkers may not be as routinely accessible. Consequently, alternative data modalities, such as urinalysis, microbiological assays (e.g., swabs), electrophysiological signals (e.g., ECG), or imaging, would be expected to yield distinct but complementary clustering structures, emphasizing organ-specific or modular and multi-scale, functional relationships.

Importantly, this suggests that the observed hematopoietic cluster should be interpreted not as a universal disease taxonomy, but as a projection of disease similarity within a blood-based feature space. Extending this framework to circulating biomarkers (e.g., liquid biopsies), multi-omics profiles, and next-generation sequencing data could enable a more integrated, multimodal disease landscape, capturing both systemic and tissue-specific dynamics. Thus, our results highlight both the strength and the inherent limitations of single-modality representations, motivating future work in multimodal integration for more comprehensive disease relationship mapping.

This interpretation is further supported by algorithmic information dynamics (AID)–based perturbation analysis, in which BDM-derived perturbation shifts identify a small subset of hematologic disease signatures whose perturbation in silico induces disproportionately large changes in the global information structure, consistent with causal influence and emergent self-organization (Figure 10; Figure S3; Table S6). By approximating generative mechanisms via minimal description length (Kolmogorov complexity) under network perturbation, AID identifies structurally emergent features beyond statistical correlation. Complementary machine learning analyses, including Random Forest feature importance (Figures 6 and 8), independently converge on the same hematologic signatures (e.g., neutrophils, RBC indices, platelets), reinforcing that the observed clustering reflects reproducible, biologically meaningful patterns rather than arbitrary feature selection. Thus, complexity metrics further demonstrate that these findings extend from descriptive clustering to causal, generative, multi-method–validated disease signatures and emergent patterns in silico.

Further, as a sensitivity analysis, we SVHC alongside effect size–based feature validation, both of which confirmed the robustness and reproducibility of the observed clustering structure. SVHC demonstrated that clusters are statistically stable under bootstrap resampling with FDR control, supporting a non-arbitrary, biologically grounded organization of disease phenotypes. These findings are consistent with an attractor-like network physiology, in which hematologic states self-organize into hierarchical, modular groupings within a shared phenotypic space.

The validated clusters correspond to coherent blocks along the diagonal of the correlation matrix, indicating strong intra-cluster similarity and a structured, multiscale organization comprising both fine-grained subclusters and higher-order groupings. Complementary effect size analysis (Supplementary Tables S2–S3) further supported this structure, revealing moderate-to-large and statistically significant differences across disease systems. Hematological profiles exhibited consistent effect size deviations across multiple feature domains (Hedges' g up to ~0.8, $q < 0.05$ after correction), whereas endocrine/metabolic and cardiovascular profiles showed comparatively attenuated or opposing shifts.

The recurrence of the same features across multiple timepoints indicates temporally stable disease signatures rather than isolated single-timepoint deviations. These effects were driven by core hematologic indices, including RBC, neutrophils, and MCV, supporting a modular and biologically meaningful organization driven by stable longitudinal trajectories. Together, these complementary results indicate that clustering is driven by reproducible system-specific hematologic patterns rather than noise fluctuations, reinforcing the interpretation of disease states as structured, attractor-like regions within a shared state-space.

These results support the notion that hematopoietic clustering reflects both methodological robustness and biological plausibility, establishing a benchmark against which the more complex patterns of non-hematopoietic diseases can be interpreted using complex systems methods.

**Heterogeneity of Non-Hematopoietic Systems**

In contrast to the coherence of hematopoietic disorders, non-hematopoietic disease categories exhibited marked heterogeneity across all analyses. In the phylogenetic tree and correlation-based clustering, most endocrine, metabolic, cardiovascular, gastrointestinal, respiratory, and nervous system diseases were distributed across a large composite cluster, lacking the compact substructures observed in hematopoietic profiles. Dimensionality reduction reinforced this dispersion: PCA captured global variance predominantly along a single axis without resolving system-level structure, while UMAP positioned non-hematopoietic conditions along a continuous manifold with extensive category overlap. Although enrichment analysis showed that these diseases contributed to immune and inflammatory pathways, their analyte profiles converged on broad systemic mechanisms rather than system-specific signatures. Several biological and methodological factors likely explain this pattern. Many non-hematopoietic conditions exert only indirect effects on circulating analytes, with endocrine disorders such as diabetes and thyroid disease frequently coexisting and disrupting shared metabolic pathways with highly variable manifestations [29], and cardiovascular or gastrointestinal diseases presenting secondary hematological changes (e.g., anemia of chronic disease, inflammatory responses, electrolyte shifts) that overlap extensively across systems [30]. These shared downstream phenotypes, compounded by syndromic complexity and comorbidities, dilute the specificity of blood-based signatures. Methodologically, Pearson correlation and PCA emphasize global variance structures, masking finer-grained distinctions, while UMAP reveals gradients rather than discrete partitions, and Random Forest prioritizes hematology-related features, further reflecting the lack of dominant analyte drivers in non-hematopoietic systems. This dispersion indicates that blood-analyte signatures alone are insufficient for stable categorization outside hematopoietic conditions, highlighting both the interpretive limitations of analyte-based clustering and the need for integrative multimodal frameworks to resolve cross-system biological overlaps.

**Inflammation**

Cluster 9, which encompassed most non-hematopoietic diseases alongside a subset of hematopoietic and “Unknown” profiles, emerged as the most heterogeneous grouping in the correlation-based clustering. Despite its size and diversity, enrichment analyses provided clear evidence that this cluster is unified by a series of shared immune and inflammatory mechanisms. KEGG pathway analysis identified highly significant enrichment for cytokine–cytokine receptor interaction, Th17 cell differentiation, AGE–RAGE signaling in diabetic complications, rheumatoid arthritis, inflammatory bowel disease, and infection-related pathways including leishmaniasis and malaria. The q-values for these pathways were extremely low (spanning approximately $4.3 \times 10^{-15}$ to $9.4 \times 10^{-23}$), underscoring the robustness of the enrichment. Notably, these processes encompassed both broad immunological signaling axes and clinically recognized disease categories, indicating that Cluster 9 is not simply a residual grouping of dissimilar conditions, but one that reflects convergence on fundamental immune–inflammatory biology.

Network physiology-based visualizations further refined this interpretation. The KEGG cnetplot revealed that multiple pathways within Cluster 9 were connected through overlapping sets of hub genes, particularly pro-inflammatory cytokines and transcriptional regulators, which bridged autoimmune diseases, infection-driven processes, and metabolic complications. The emapplot corroborated these findings by highlighting functional modules: one linking leishmaniasis, tuberculosis, and toxoplasmosis, another connecting Th17 differentiation, rheumatoid arthritis, and inflammatory bowel disease, and a third coupling cytokine–cytokine receptor interaction with viral protein interaction with cytokine signaling. These modules illustrate how disparate clinical diagnoses can converge on shared gene regularity networks, explaining why such diverse entities are grouped together in correlation space.

Reactome analysis provided complementary evidence, emphasizing interleukin signaling and platelet/ECM biology as key bridging processes. Interleukin pathways (Interleukin-4/13 and Interleukin-10 signaling) clustered tightly, reflecting central cytokine cascades in autoimmune and inflammatory contexts, while platelet activation, degranulation, and calcium-mediated responses formed a parallel submodule linked to extracellular matrix organization and degradation. These findings indicate that although Cluster 9 is clinically heterogeneous, it is biologically coherent at the mechanistic level, converging on shared immune, inflammatory, and hemostatic pathways. This demonstrates that analyte-based similarity can uncover cross-cutting biological processes that transcend traditional system-based classifications and may inform refinements to disease taxonomy.

**Limitations and System-Level Interpretation**

The following subsections are retained verbatim, now grouped under a single interpretive heading to reduce structural fragmentation while preserving all original content exactly as provided. While the analyses presented here provide a coherent account of disease (dis)similarity based on blood-analyte signatures, several methodological considerations and limitations warrant careful discussion.

Another important limitation of this study is the use of synthetic disease signatures derived from curated medical knowledge rather than real patient-level data. While this approach enables controlled exploration of computational frameworks and mechanistic structure, it does not fully capture the complexity, measurement noise, and population heterogeneity inherent to patient-derived clinical datasets. Consequently, the present findings should be interpreted as demonstrating the feasibility and internal consistency of analyte-based disease modeling in silico using network physiology methods, rather than definitive biological conclusions. Future validation using real-world clinical cohorts will be essential to confirm the robustness and translational applicability of these network patterns.

**Choice of similarity metric**

Disease–disease similarity was quantified using Pearson correlation, which captures linear associations between analyte profiles. Although widely used and computationally efficient, this approach may underestimate relationships that are non-linear or driven by rare analytes. Alternative metrics, such as Spearman correlation [31], cosine similarity [32], or distance measures more robust to outliers, might reveal additional structure. Moreover, converting correlation to distance through $1 - \rho$ assumes symmetry and equal weight across all analytes, potentially obscuring clinically meaningful heterogeneity.

**Clustering and dimensionality reduction**

The hierarchical clustering with UPGMA provided an interpretable dendrogram and was useful for partitioning into clusters. However, cluster membership is sensitive to the choice of cut threshold, as demonstrated by the large and heterogeneous Cluster 9. Similarly, PCA captured global variance but failed to separate systemic categories beyond hematopoietic diseases, reflecting the limitations of linear projections [26]. UMAP recovered more nuanced local structures but is inherently stochastic and parameter-dependent [27]; alternative embeddings or repeated runs might yield slightly different layouts. These issues highlight that observed separations are robust only when consistent across multiple methods, as was the case for hematopoietic clustering.

**Feature attribution**

Random Forest classification identified neutrophils, red blood cell indices, haemoglobin, and platelets as key features. While these results align with biological plausibility, tree-based feature importance is influenced by collinearity among analytes and repeated measurements across timepoints. Importances therefore indicate relative contribution within the model, rather than causal effects.

The findings parallel, yet diverge from organ-based ICD taxonomies, suggesting that inflammatory and immune-network topology may provide a complementary axis of biological disease classification.

**Sensitivity and Robustness of Feature Space Organization**

Given that manifold and representation learning approaches in **network physiology** are inherently sensitive to methodological variability and noise, we performed post-hoc complex systems theory-driven analyses to assess robustness and reproducibility of the observed network patterns. PCA structure remained consistent across imputation strategies, indicating that missing data handling does not bias global variance patterns. Similarly, UMAP embeddings were stable across random seeds, with high Procrustes similarity and preserved distance relationships (Tables S5–S6), confirming that clustering is not driven by stochastic initialization but rather reflects self-organized structure.

Noise perturbation analysis further demonstrated that both the time-series and learned global feature spaces are resilient to low-to-moderate Gaussian noise, with minimal disruption to variance structure and embedding topology (Figure 9; Tables 2–4). Importantly, imputation strategy did not alter variance structure in either feature space representations (both, global and time-series). However, a key distinction emerges: the time-series feature space is highly compressible and low-dimensional, whereas the learned global feature space captures more distributed and informative variation in the clustering patterns by disease states. Complementary SVHC validation, ML classification, and AID-based feature perturbation analyses provide convergent evidence that the observed clustering pattern is reproducible and not an artifact of clustering methodology or data construction.

**Attractor Inference and State-Space Geometry**

Analysis of the 55-feature longitudinal time-series space revealed a self-organized clustering pattern, with clustering consistently optimal at $k = 2$ and reduced cohesion at higher k, indicating a dominant two-state structure. PCA further showed that variance is overwhelmingly captured by a single component (PC1 $\approx$ 0.93), with >99% explained by the first five PCA components. These findings suggest that temporal trajectories are highly compressible and governed by constrained transitions among the disease states' hematological features.

This supports an attractor-like interpretation in routine blood data, where disease trajectories converge onto a limited number of stable hematological states. In contrast, the learned global feature space exhibits a more distributed variance profile (PC1 $\approx$ 0.38), revealing a higher-dimensional manifold geometry with richer multiscale separation between disease spatial profiles. Thus, temporal dynamics are compressible and attractor-like, whereas the global (spatial) feature space encodes complex geometric structure.

Furthermore, nonlinear manifold analysis reinforces this attractor-like interpretation. UMAP

stability across seeds demonstrates that both spaces preserve highly conserved relational structure, indicating that the topology of the disease state space is intrinsic and robust (Supplementary Info: Table S5-S6, Figure S2). AID and ML analyses further support this interpretation, identifying consistent feature patterns that align with this manifold geometry. As shown in Figure 10, AID-based perturbation analysis supports an attractor-like organization of the disease state-space, where a small subset of hematologic features exerts control over global system structure. The dominance of BDM over entropy and compression metrics highlights its sensitivity to algorithmic, rather than purely statistical, complexity. The recurrence of these features across timepoints further reinforces the presence of network regulatory/control points. Within the broader context of network physiology and dynamical systems interpretation, these findings suggest that disease states mapped by blood-derived features occupy structured regions of a shared state-space, where stable, low-dimensional temporal trajectories are embedded within a more complex global geometric landscape.

## Future Directions

The current analysis was based on a curated repository of 103 disease profiles, each derived from knowledge-based synthetic blood-analyte signatures. The disease profiles were curated as part of a synthetic digital signature repository, each containing selected analytes measured at multiple timepoints. While this dataset enabled proof-of-concept analyses, the size and diversity remain limited. Future studies should incorporate clinically derived data beyond synthetic, larger and more heterogeneous cohorts ideally spanning multiple populations, age groups, and clinical settings. Such expansion would increase statistical power, reduce biases, and enable stratified analyses that account for demographic and comorbidity effects. Linking profiles to longitudinal data would further allow investigation of disease progression and treatment response.

Blood-analyte profiles capture only one dimension of disease biology. Integrative frameworks that combine hematology with symptoms, and imaging should be explored. Multi-omic integration would help disentangle heterogeneous systems such as endocrine or gastrointestinal diseases, where blood indices provide only indirect readouts. Incorporating electronic health records and patient-reported outcomes could bridge molecular signatures with clinical trajectories, enhancing translational relevance.

Ultimately, future work should explore how analyte-based clustering can contribute to a hybrid taxonomy that integrates ICD-based classification with cross-system mechanistic axes. This may enable more biologically meaningful groupings, support early disease interception, and guide mechanism-based treatment allocation. Moving from proof-of-concept toward clinical-grade classification will require iterative refinement between computational analyses, mechanistic interpretation, and translational implementation. Future studies should also incorporate causal-mechanistic frameworks capable of revealing how minimal perturbations to blood-analyte signatures propagate through disease clusters and shared inflammatory axes. Approaches grounded in algorithmic information dynamics (AID**)** offer such a way to infer generative structure, identify causal drivers of cluster formation, and model counterfactual intervention strategies directly from digital blood twins [37–40]. Our findings support the AID framework within network physiology, where perturbation analysis reveals a small subset of structurally influential features that disproportionately shape the global information structure, consistent with control-like behavior in an attractor landscape. Integrating these causal tools with larger, real-world datasets may help refine mechanistic disease taxonomy, expose latent confounding pathways, and enhance the interpretability of analyte-based disease stratification.

## Conclusions

This study demonstrates that blood-derived digital signatures can recover highly coherent disease clusters for hematopoietic conditions while simultaneously revealing a large, mechanistically convergent cluster of systemic disorders driven by immunological and inflammatory pathways. Rather than replicating ICD-style taxonomies, hematology-based similarity exposes confounding factors and a deeper mechanistic substrate, one in which cytokine signaling, interleukin networks, platelet activation, and extracellular matrix remodeling form shared axes that unify clinically disparate diseases.

In translational and systems medicine, this suggests that routine laboratory data, already embedded in healthcare workflows, could serve as the basis for a complementary mechanistic disease ontology. Such a hybrid taxonomy, grounded partly in analyte biology and partly in clinical categorization, would support earlier risk stratification, more granular comorbidity mapping, and biologically informed patient grouping. Importantly, this network physiology framework does not require new tests, technologies, or invasive sampling; it reuses existing clinical data to surface latent biological structure.

By linking hematologic coherence with immune–inflammatory convergence across complex systems, this work outlines a pathway toward precision medicine that is computationally accessible and clinically scalable. As future validation studies expand the dataset, integrate multi-modal signals, and evaluate prognostic value, analyte-based digital twins may evolve into real-time risk stratification tools capable of informing continuous disease monitoring, mechanistic (causal) discovery, and precision treatment allocation in patient-centered care.

**Supplementary Material**

Table S1: Disease profiles grouped by hierarchical clustering (cut at distance = 0.02). Profile IDs are omitted for clarity; only disease names are shown.

| *Cluster* | *Number of Diseases* | *Disease Names* |
|---|---|---|
| *1* | 1 | Acute Myeloid Leukemia (AML) |
| *2* | 1 | Pancytopenia |
| *3* | 1 | AML2 |
| *4* | 1 | Lymphoma2 |
| *5* | 2 | Viral Lower Respiratory Tract Infection; Bacterial Pneumonia |
| *6* | 6 | Myeloma; Acquired Lymphopenia; CKD v2; Aortic Dissection; Meningitis v2; Febrile Neutropenia |
| *7* | 1 | Myelodysplastic Syndrome |
| *8* | 1 | Pancytopenia v2 |
| *9* | 80 | Microcytic Anaemia; Microcytic Anaemia v2; Normocytic Anaemia; Mixed Micro/Macrocytic Anaemia; Polycythaemia; Essential Thrombocythaemia; Acute Haemolytic Anaemia; Myelofibrosis; Drug-induced Neutropenia; Macrocytic Anaemia; Chronic Myeloid Leukemia; Chronic Lymphocytic Leukemia; Acute Blood Loss (Upper GI); Acute Blood Loss (Lower GI); Chronic Blood Loss; Allergic Reaction; Acute IBD Flare; Type 1 Diabetes; Type 2 Diabetes; Diabetic Ketoacidosis; Hyperthyroidism; Hypothyroidism; Hyper-/Hypokalaemia; Hyper-/Hypocalcaemia; Acute Kidney Injury (v1–v3); Chronic Kidney Disease; Hyponatraemia (Hypo-/Hyper-/Normovolaemic); Hypernatraemia; Adrenal Insufficiency; Hypophosphataemia; Hypomagnesaemia; Myopericarditis; Pericarditis; Infective Endocarditis; Myocardial Infarction (NSTEMI/STEMI); Heart Failure; Pulmonary Embolism; Deep Vein Thrombosis; Urinary Tract Infection; Pyelonephritis; Bacterial Pharyngitis; COPD; Asthma; Bronchiectasis; Empyema; Gastroenteritis; Hepatitis (v1/v2); Gilbert Syndrome; Cholangitis; Biliary Colic; Cellulitis; Osteomyelitis; Meningitis; Cholangiopathy; Atypical Pneumonia; Acute Pancreatitis; Chronic Diarrhoea (Coeliac); Diverticulitis; Gout; Giant Cell Arteritis; Rheumatoid Arthritis; Eosinophilic Vasculitis; Polymyalgia Rheumatica |
| *10* | 3 | CML2; CLL2; Lymphoma |
| *11* | 1 | ALL2 |
| *12* | 1 | Heparin-Induced Thrombocytopenia (HIT) |
| *13* | 1 | Immune Thrombocytopenia (ITP) |
| *14* | 1 | Acute Lymphoblastic Leukemia (ALL) |
| *15* | 1 | Thrombotic Thrombocytopenic Purpura (TTP) |
| *16* | 1 | Chronic Liver Disease |

| *Contrast* | *Feature* | *Hedges' g* | *p-value* | *q-value (BH)* |
|---|---|---|---|---|
| *Infection vs Rest* | MCV (fL)_t1 | -0.586 | 6.82E-07 | 3.70E-05 |
| *Infection vs Rest* | MCV (fL)_t3 | -0.632 | 7.98E-06 | 2.19E-04 |
| *Rheumatological vs Rest* | Segmented neutrophils (1000/µL)_t3 | -0.541 | 1.07E-05 | 5.89E-04 |
| *Infection vs Rest* | MCV (fL)_t5 | -0.488 | 3.77E-04 | 6.91E-03 |
| *Rheumatological vs Rest* | WBC (1000/µL)_t3 | -0.511 | 2.82E-04 | 7.75E-03 |
| *Infection vs Rest* | WBC (1000/µL)_t2 | 0.513 | 1.31E-03 | 1.80E-02 |
| *Haematological vs Rest* | Segmented neutrophils (1000/µL)_t5 | -0.735 | 1.34E-04 | 2.26E-02 |
| *Haematological vs Rest* | Segmented neutrophils (1000/µL)_t4 | -0.727 | 7.44E-04 | 2.26E-02 |
| *Rheumatological vs Rest* | WBC (1000/µL)_t2 | -0.387 | 1.35E-03 | 2.47E-02 |
| *Rheumatological vs Rest* | Segmented neutrophils (1000/µL)_t2 | -0.46 | 2.08E-03 | 2.86E-02 |
| *Infection vs Rest* | WBC (1000/µL)_t3 | 0.715 | 3.78E-03 | 2.97E-02 |
| *Infection vs Rest* | MCV (fL)_t4 | -0.511 | 3.76E-03 | 2.97E-02 |
| *Infection vs Rest* | WBC (1000/µL)_t1 | 0.389 | 3.39E-03 | 2.97E-02 |
| *Infection vs Rest* | WBC (1000/µL)_t4 | 0.705 | 4.53E-03 | 3.12E-02 |
| *Infection vs Rest* | WBC (1000/µL)_t5 | 0.823 | 6.01E-03 | 3.30E-02 |
| *Infection vs Rest* | MCV (fL)_t2 | -0.501 | 5.64E-03 | 3.30E-02 |
| *Haematological vs Rest* | RBC (million/µL)_t4 | -0.844 | 2.05E-05 | 4.71E-02 |

**Table S2. Top 20 time-series feature-level effect sizes across disease-system contrasts**

Feature-level differences between disease systems and the remainder of the dataset were quantified using Hedges' g across longitudinal time-series measurements (t1 to t5). Statistical significance was assessed using Welch's t-test or Mann–Whitney U test depending on normality (Shapiro–Wilk test), with Benjamini–Hochberg correction applied for multiple comparisons. Features are ranked by adjusted q-values and effect size magnitude. Dominant signals arise from red blood cells (RBC), neutrophils, mean corpuscular volume (MCV), and white blood cell count (WBC), demonstrating consistent discriminative patterns across timepoints.

SVHC hierarchical clustering confirmed that the time-series global disease-profile structure is statistically robust, identifying 25 validated clusters ($p < 0.01$) corresponding to statistically robust intra-cluster correlation blocks. Complementary effect size analysis reveals that these clusters are driven by coherent, system-specific hematologic trajectories over time, moderate-to-large, system-specific hematologic separations (Hedges' g up to ~0.8, $q < 0.05$ after correction), with consistent signals across longitudinal timepoints.

Importantly, the recurrence of the same features (e.g., MCV, WBC, neutrophils) across multiple timepoints (t1–t5) indicates that disease differentiation is not static but reflects stable longitudinal dynamics, supporting a time-resolved, modular organization of the digital blood phenotypic space. Infection profiles exhibit consistent leukocytic elevation patterns, rheumatologic conditions show coordinated suppression or redistribution of immune cell populations, and hematological diseases demonstrate the largest magnitude deviations in erythroid and myeloid compartments. These findings support a complex system, temporally structured (i.e., attractor-like) architecture in which disease states are defined by reproducible trajectories of hematological patterns.

| *system_a* | *system_b* | *pearson_r* | *pearson_p* | *spearman_rho* | *spearman_p* |
|---|---|---|---|---|---|
| *Other* | Cardiac | 0.999729 | 2.98E-88 | 0.986265 | 3.93E-43 |
| *Other* | Other | 0.99961 | 4.84E-84 | 0.989242 | 6.28E-46 |
| *Haematological* | Infection | 0.999597 | 1.12E-83 | 0.993759 | 3.60E-52 |
| *Other* | GI_Hepatic | 0.99952 | 1.19E-81 | 0.985727 | 1.08E-42 |
| *Haematological* | Other | 0.999457 | 3.07E-80 | 0.970158 | 2.74E-34 |
| *Other* | Other | 0.999442 | 6.17E-80 | 0.984958 | 4.30E-42 |
| *Other* | Other | 0.999407 | 3.12E-79 | 0.977414 | 1.86E-37 |
| *Other* | Cardiac | 0.99936 | 2.42E-78 | 0.97652 | 5.16E-37 |
| *Haematological* | Other | 0.99935 | 3.63E-78 | 0.960414 | 4.33E-31 |
| *Other* | Other | 0.999347 | 3.98E-78 | 0.981817 | 6.29E-40 |
| *Other* | Other | 0.999318 | 1.28E-77 | 0.983634 | 3.95E-41 |
| *Cardiac* | GI_Hepatic | 0.999312 | 1.64E-77 | 0.977081 | 2.74E-37 |
| *Haematological* | Infection | 0.99922 | 4.43E-76 | 0.967898 | 1.84E-33 |
| *Other* | Other | 0.999218 | 4.77E-76 | 0.946801 | 9.21E-28 |
| *Other* | Cardiac | 0.999215 | 5.39E-76 | 0.988943 | 1.30E-45 |
| *Haematological* | Other | 0.999209 | 6.57E-76 | 0.968878 | 8.21E-34 |
| *Other* | GI_Hepatic | 0.999209 | 6.57E-76 | 0.945036 | 2.14E-27 |
| *Haematological* | Respiratory | 0.999189 | 1.26E-75 | 0.929186 | 1.44E-24 |
| *Other* | Other | 0.999186 | 1.38E-75 | 0.946073 | 1.31E-27 |
| *Respiratory* | GI_Hepatic | 0.999158 | 3.37E-75 | 0.985181 | 2.90E-42 |

**Table S3. Pairwise disease profile correlations across systems**/ Top-ranking pairwise correlations between disease profiles derived from the 55-analyte time-series feature space. Pearson (linear) and Spearman (rank-based) correlation coefficients with corresponding p-values are shown for representative high-similarity profile pairs across physiological systems.
These results demonstrate near-universal high concordance between disease profiles (Pearson $r \approx 0.999$), indicating strong global similarity in analyte patterns across systems, with consistently significant monotonic relationships (Spearman $\rho$ up to ~0.99), supporting a shared underlying structure in the digital blood phenotypic space. The Pearson confidence intervals were also within the 0.99 to 0.999 range.

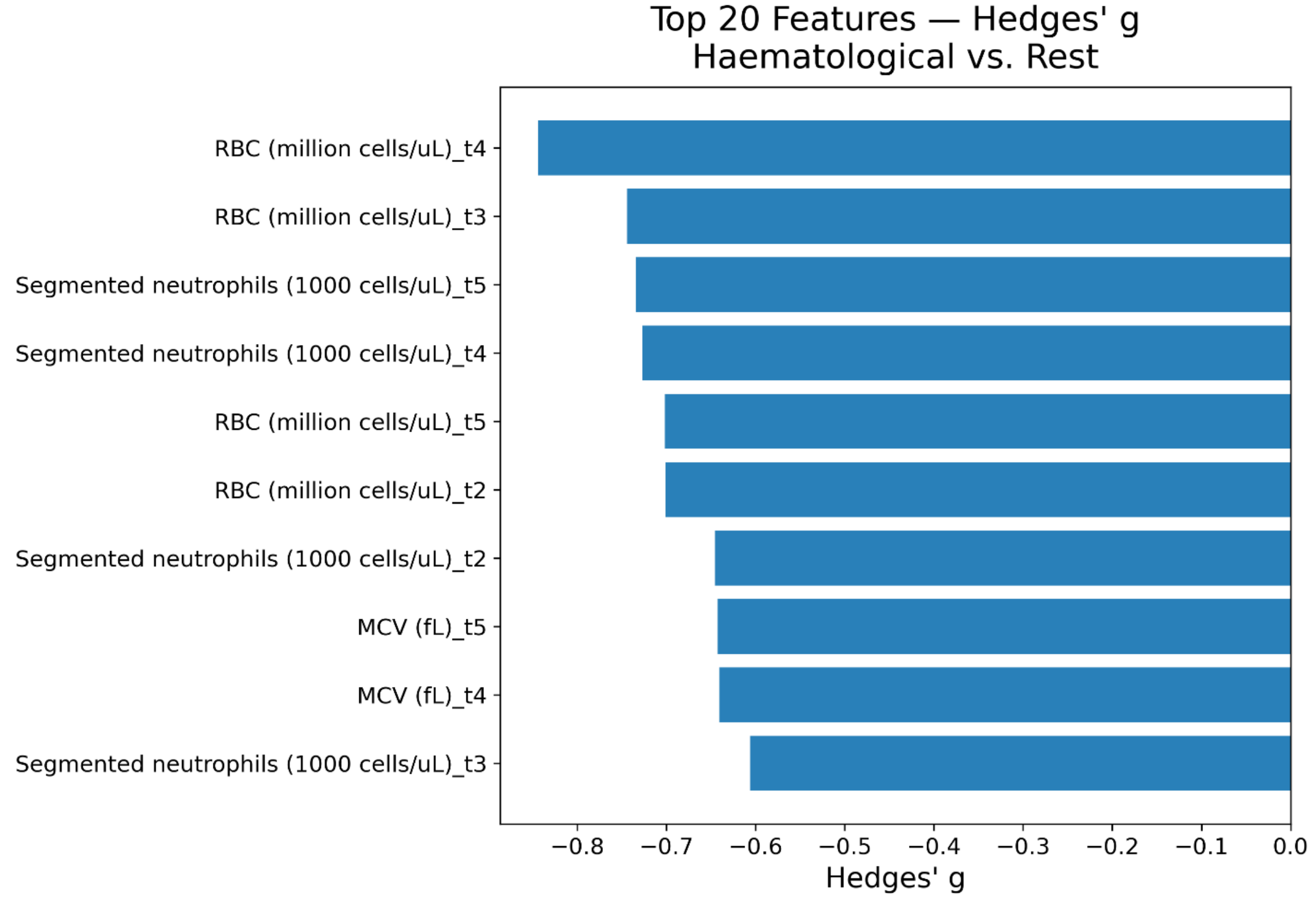

**Figure S1. Top 20 time-series feature effect sizes (Hedges' g): Haematological vs. Rest**
Top 20 discriminative longitudinal features ranked by Hedges' g comparing haematological disease profiles to all other systems. Effect sizes are consistently negative, reflecting reduced values in haematological conditions across repeated timepoints (t2–t5). Dominant signals arise from erythroid (RBC), leukocyte (segmented neutrophils), and red cell morphology (MCV) indices, demonstrating stable, time-resolved separation of haematological profiles.

**Feature Importance Analysis using Machine Learning Multiclass Validation**

The analysis identified a consistent set of highly robust and discriminative features, with longitudinal diagnostic and monitoring signatures (e.g., *fbc_dx_longitudinal_with_signature*, *fbc_dx_single_with_signature*, and *monitor_longitudinal_without_signature*) ranking highest across all importance metrics and exhibiting near-perfect top-feature frequency (≈1.0). These features also demonstrated the lowest mean ranks and minimal variability, indicating strong stability across bootstrap resampling. In contrast, system-level categorical features (e.g., cardiovascular, neuro, thrombosis) showed minimal importance and low selection frequency, suggesting that classification performance is driven primarily by structured hematologic signatures rather than broad disease labels. The convergence of Gini, permutation, SHAP, and bootstrap metrics further confirms that feature importance is robust and not ML method dependent. Collectively, these results support a stable pattern, i.e., a reproducible feature hierarchy underlying disease separability.

| feature | gini_importance | perm_mean | perm_std | shap_mean_abs | bootstrap_mean_gini | bootstrap_sd_gini | mean_rank | sd_rank | top20_frequency |
|---|---|---|---|---|---|---|---|---|---|
| cardiovascular | 0.0008 | 0.0000 | 0.0000 | 0.0011 | 0.0015 | 0.0012 | 18.7200 | 2.1622 | 0.8200 |
| endocrine_metabolic | 0.0004 | 0.0000 | 0.0000 | 0.0003 | 0.0006 | 0.0006 | 21.5850 | 2.1178 | 0.2600 |
| fbc_dx_longitudinal_with_signature | 0.1298 | 0.0052 | 0.0048 | 0.0684 | 0.1491 | 0.0245 | 2.4800 | 0.7698 | 1.0000 |
| fbc_dx_longitudinal_without_signature | 0.0036 | 0.0000 | 0.0000 | 0.0018 | 0.0055 | 0.0034 | 15.2400 | 1.5112 | 0.9850 |
| fbc_dx_single_with_signature | 0.1628 | 0.0052 | 0.0048 | 0.0841 | 0.1446 | 0.0247 | 2.6950 | 0.7380 | 1.0000 |
| fbcplus_dx_longitudinal_with_signature | 0.0473 | 0.0025 | 0.0043 | 0.0312 | 0.0464 | 0.0122 | 8.2450 | 1.6024 | 1.0000 |
| fbcplus_dx_longitudinal_without_signature | 0.0452 | 0.0025 | 0.0043 | 0.0284 | 0.0464 | 0.0123 | 8.2250 | 1.6147 | 1.0000 |
| fbcplus_dx_single_with_signature | 0.0604 | 0.0025 | 0.0043 | 0.0382 | 0.0620 | 0.0162 | 6.0750 | 1.3596 | 1.0000 |
| gi_hepatobiliary | 0.0006 | 0.0000 | 0.0000 | 0.0006 | 0.0011 | 0.0006 | 19.6150 | 1.6705 | 0.7150 |
| hematology_anemia | 0.0015 | 0.0000 | 0.0000 | 0.0028 | 0.0032 | 0.0026 | 16.6300 | 1.7402 | 0.9750 |
| hematology_bleeding | 0.0227 | 0.0000 | 0.0000 | 0.0156 | 0.0278 | 0.0190 | 10.8050 | 2.5533 | 1.0000 |
| hematology_leukemia_lymphoma | 0.0001 | 0.0000 | 0.0000 | 0.0001 | 0.0003 | 0.0003 | 22.8600 | 1.7044 | 0.0950 |
| hematology_platelet | 0.0000 | 0.0000 | 0.0000 | 0.0001 | 0.0004 | 0.0005 | 22.5500 | 1.9407 | 0.1450 |
| hematology_splenomegaly | 0.0242 | 0.0000 | 0.0000 | 0.0161 | 0.0270 | 0.0170 | 10.8800 | 2.3113 | 1.0000 |
| infection_inflammation | 0.0003 | 0.0000 | 0.0000 | 0.0001 | 0.0006 | 0.0004 | 21.2900 | 1.5647 | 0.2850 |
| monitor_longitudinal_with_signature | 0.0292 | 0.0000 | 0.0000 | 0.0175 | 0.0346 | 0.0104 | 9.7850 | 1.6099 | 1.0000 |
| monitor_longitudinal_without_signature | 0.2626 | 0.0000 | 0.0000 | 0.1320 | 0.2150 | 0.0291 | 1.1500 | 0.4782 | 1.0000 |
| monitor_single_with_signature | 0.0102 | 0.0000 | 0.0000 | 0.0049 | 0.0116 | 0.0054 | 13.1550 | 1.1781 | 1.0000 |
| neuro | 0.0008 | 0.0000 | 0.0000 | 0.0004 | 0.0009 | 0.0004 | 20.1000 | 1.5977 | 0.6400 |
| renal_urinary | 0.0027 | 0.0000 | 0.0000 | 0.0023 | 0.0032 | 0.0020 | 16.5800 | 1.6051 | 0.9850 |
| respiratory | 0.0087 | 0.0000 | 0.0000 | 0.0079 | 0.0097 | 0.0067 | 13.7900 | 1.3913 | 1.0000 |
| screening_longitudinal_with_signature | 0.0465 | 0.0000 | 0.0000 | 0.0231 | 0.0532 | 0.0156 | 7.2200 | 1.8866 | 1.0000 |
| screening_longitudinal_without_signature | 0.0451 | 0.0000 | 0.0000 | 0.0214 | 0.0519 | 0.0144 | 7.4100 | 1.9341 | 1.0000 |
| screening_single_with_signature | 0.0945 | 0.0000 | 0.0000 | 0.0490 | 0.1032 | 0.0252 | 3.8600 | 0.8913 | 1.0000 |
| thrombosis | 0.0000 | 0.0000 | 0.0000 | 0.0000 | 0.0002 | 0.0005 | 23.9200 | 2.0918 | 0.0950 |

**Table S4. Integrated feature importance and robustness analysis across machine learning and statistical validation methods.** Feature importance was evaluated across multiple complementary methods, including Random Forest Gini importance, permutation importance (mean ± SD), SHAP mean absolute values, and bootstrap stability of Gini importance (mean ± SD). Features were further ranked using a consensus mean rank and ranking variability (SD), and their frequency of appearance in the top 20 features across bootstrap iterations was recorded. This integrated framework enables assessment of both discriminative power and robustness of feature contributions across models and resampling strategies.

| *seed* | *procrustes_similarity_to_seed0* | *distance_matrix_corr_to_seed0* |
|---|---|---|
| *1* | 0.772276 | 0.906219 |
| *2* | 0.950362 | 0.976061 |
| *3* | 0.96088 | 0.988202 |
| *4* | 0.633593 | 0.916613 |
| *5* | 0.97719 | 0.982387 |
| *6* | 0.960889 | 0.977868 |
| *7* | 0.822286 | 0.920419 |
| *8* | 0.705172 | 0.942354 |
| *9* | 0.775079 | 0.947069 |
| *10* | 0.852676 | 0.97287 |
| *11* | 0.932431 | 0.978718 |
| *12* | 0.83072 | 0.940835 |
| *13* | 0.966421 | 0.97497 |
| *14* | 0.915191 | 0.954964 |
| *15* | 0.93139 | 0.954894 |
| *16* | 0.832076 | 0.97501 |
| *17* | 0.95007 | 0.978638 |
| *18* | 0.89869 | 0.97217 |

| 19 | 0.757409 | 0.959571 |
|---|---|---|

**Table S5. UMAP Embedding Stability Across Random Seeds (Global Feature Space)**
Assessment of embedding stability across multiple random initializations (seeds 1–19) relative to a reference embedding (seed 0), computed on the global aggregated feature space derived from profile-level representations (machine learning/AID feature set) rather than time-series analytes. Stability was quantified using Procrustes similarity of UMAP embeddings and correlation of pairwise distance matrices.

High Procrustes similarity and distance matrix correlations indicate preservation of global and local structure across stochastic embedding runs. UMAP embeddings demonstrated high reproducibility across seeds, with Procrustes similarity frequently exceeding 0.90 and distance correlations consistently high (≈0.94–0.99). Despite minor variability in layout (expected for stochastic methods), the underlying relational structure between disease profiles remained stable. These results confirm that the learned disease representation in the global feature space is robust to stochastic embedding variation.

| *seed* | *procrustes_similarity_vs_seed0* | *distance_matrix_corr_vs_seed0* |
|---|---|---|
| 1 | 0.793096 | 0.991479 |
| 2 | 0.948366 | 0.996104 |
| 3 | 0.968621 | 0.99368 |
| 4 | 0.991399 | 0.997421 |
| 5 | 0.952584 | 0.996354 |
| 6 | 0.872759 | 0.941837 |
| 7 | 0.750205 | 0.984684 |
| 8 | 0.989389 | 0.998995 |
| 9 | 0.991015 | 0.997774 |
| 10 | 0.976072 | 0.99818 |
| 11 | 0.991138 | 0.997539 |
| 12 | 0.986424 | 0.998606 |
| 13 | 0.860971 | 0.984069 |
| 14 | 0.912811 | 0.995184 |
| 15 | 0.477368 | 0.982724 |
| 16 | 0.980256 | 0.997966 |
| 17 | 0.971898 | 0.993647 |
| 18 | 0.988232 | 0.998948 |
| 19 | 0.824567 | 0.94936 |

**Table S6. UMAP Embedding Stability Across Random Seeds (Time-Series Feature Space)**

Assessment of embedding stability across multiple random initializations (seeds 1–19) relative to a reference embedding (seed 0), computed on the 55-feature longitudinal time-series space. Stability was quantified using Procrustes similarity of UMAP embeddings and correlation of pairwise distance matrices.

High distance matrix correlations (≈0.94–0.999) indicate strong preservation of global pairwise relationships across stochastic embedding runs, while Procrustes similarity shows moderate-to-high variability (≈0.48–0.99), reflecting sensitivity of absolute layout to initialization. Despite this variability, the underlying relational structure between profiles remains consistent.

These results indicate that the time-series feature space exhibits stable global structure, but greater geometric variability compared to the learned global feature space, as dynamical systems, consistent with a lower-dimensional and less constrained representation.

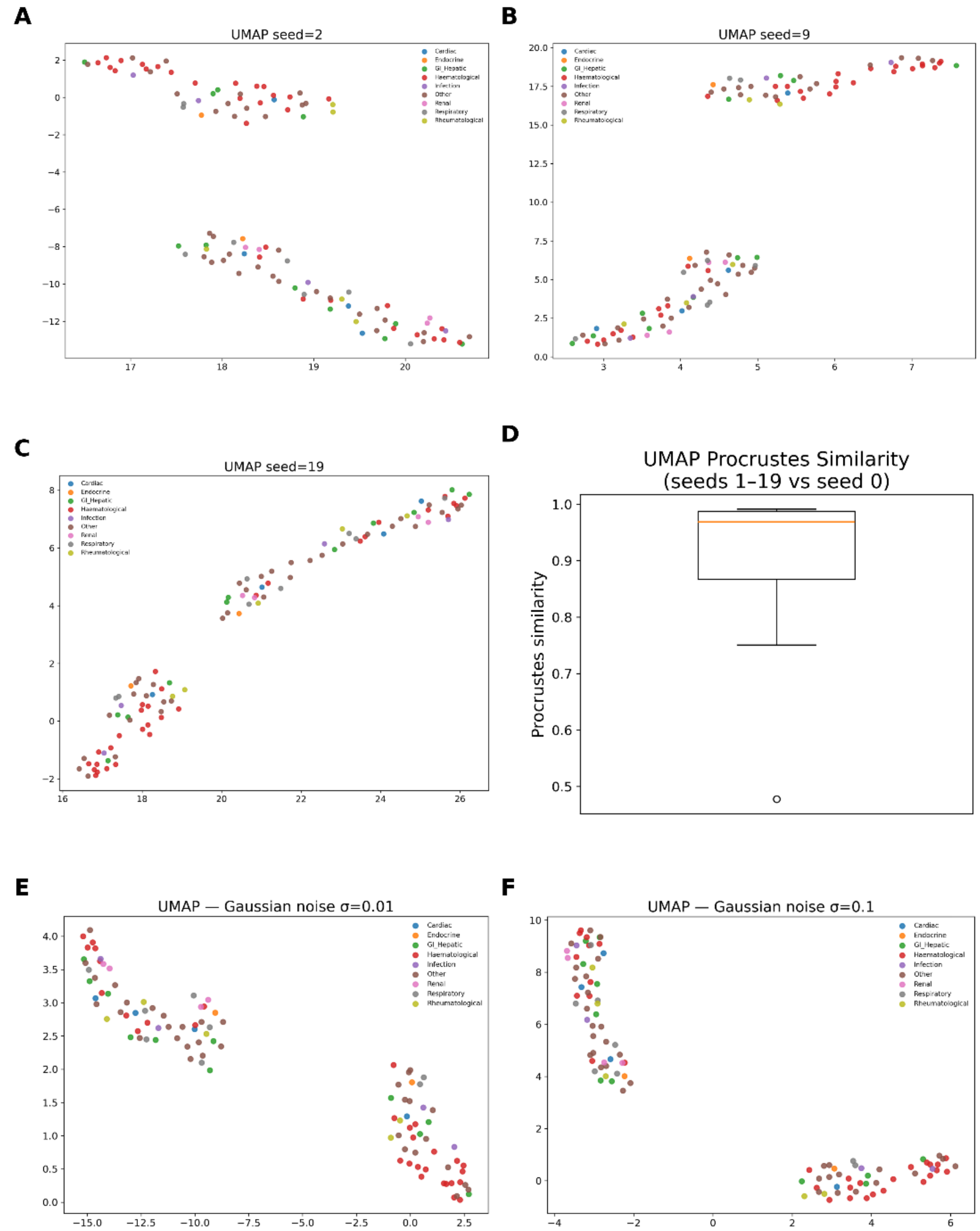

**Figure S2. UMAP stability and noise sensitivity in the 55-time-series feature space.** (A–C) UMAP embeddings of the 55-feature longitudinal time-series space across different random initializations (seeds 2, 9, and 19) demonstrate consistent global organization despite stochastic variation in layout. (D) Procrustes similarity across seeds confirms high embedding reproducibility, indicating preservation of pairwise relationships and overall structure. (E–F) UMAP embeddings under increasing Gaussian noise ($\sigma = 0.01$ and $\sigma = 0.1$) show that the dominant structural pattern remains stable under low-to-moderate perturbations, with only minor deformation at higher noise levels. Together, these results indicate that the time-series feature space captures a robust but low-dimensional structure, where global relationships are preserved across both stochastic initialization and noise perturbation.

**AID Perturbation Analysis**

Algorithmic Information Dynamics (AID) is a causal inference method in silico which combines algorithmic information theory and graph perturbation analysis to investigate the dynamics and complexity of complex networks/systems. AID asks how much the algorithmic information content of an underlying topology or representation changes when a network feature (i.e., an analyte, node, or edge) is perturbed in the disease-feature matrix. Features with the largest perturbation shifts among its complexity metrics are interpreted as the most structurally and causally influential.

In our data, the strongest and most recurrent shifts are concentrated in a small group of features, especially screening- and diagnosis-related longitudinal signatures and selected FBC-derived signatures. This further validates our findings that a limited set of features appears to carry disproportionate causal or control-like influence over the clustering pattern (representation learning). Thus, AID perturbation analysis and model-agnostic robustness complements the supervised network physiology and feature-importance results in the Main Text by highlighting structurally sensitive features across multiple complexity measures. Each feature was binarized across multiple thresholds ($t = 0.2, 0.35, 0.5, 0.65, 0.8$), and perturbation shifts were quantified using various complexity metrics as shown in Figure S3.

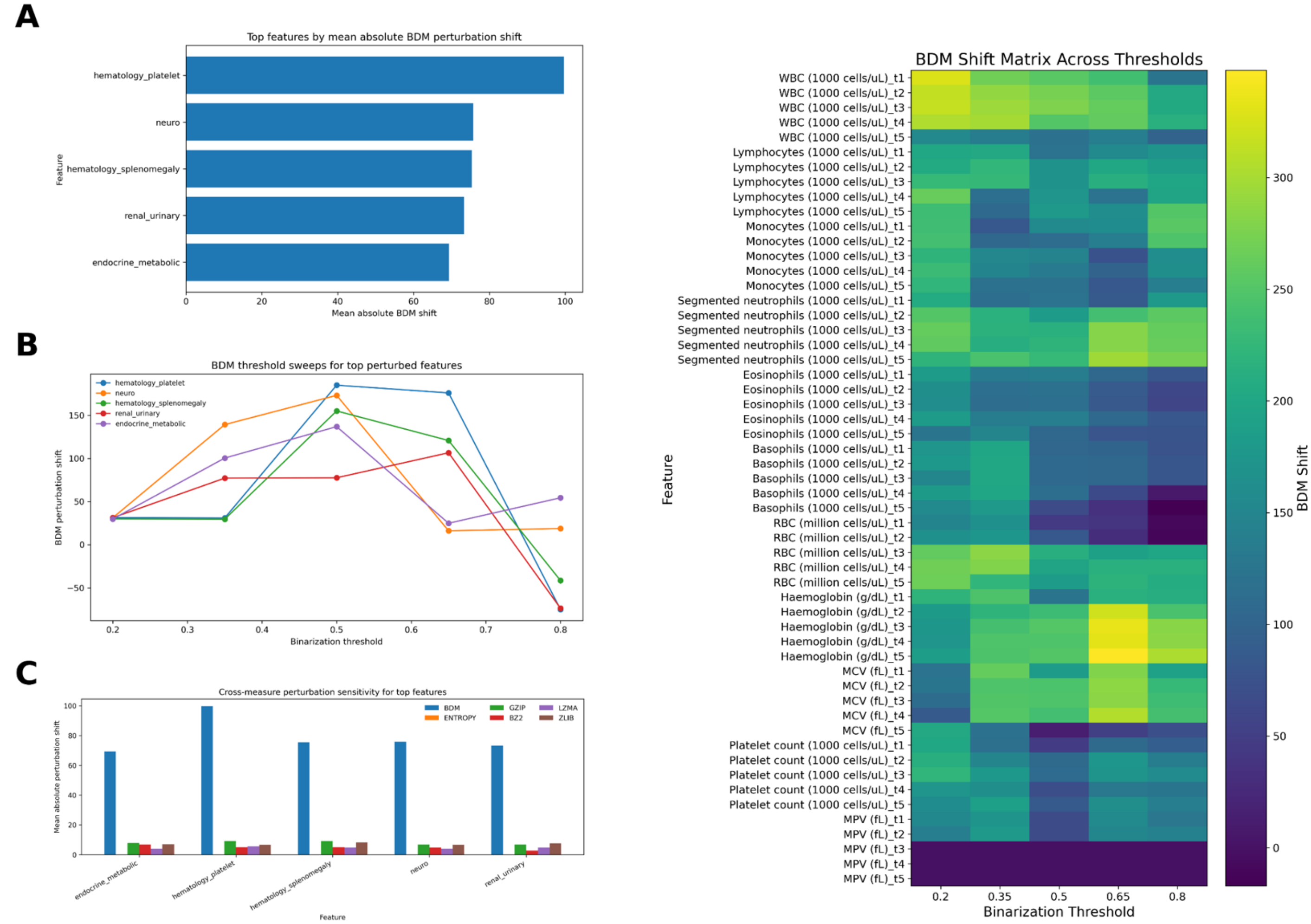

**Figure S3 Top. Algorithmic Information Dynamics (AID) perturbation analysis of feature sensitivity**. (A) Top features ranked by mean absolute BDM perturbation shift. (B) Threshold-dependent BDM perturbation profiles across binarization levels. (C) Cross-measure comparison of perturbation sensitivity using BDM, entropy, and compression-based metrics. These analyses identify structurally influential features within the hematologic feature space. **Bottom. BDM Shift Matrix Across Thresholds for Full Time-Series Disease-Feature Matrix.** Heatmap of BDM compelxity shifts for all analytes across a range of binarization thresholds (0.2–0.8) in the 55-feature time-series disease matrix. A matrix is a 2D representation of a graph network. Rows represent longitudinal analytes (e.g., WBC, haemoglobin, RBC, neutrophils), and columns represent threshold levels used in AID perturbation analysis. Higher BDM shifts indicate greater structural influence on the global information landscape. The matrix reveals consistent, high-magnitude perturbations in key hematologic features across multiple timepoints, supporting their role as dominant contributors to the system’s underlying attractor-like structure